\documentclass[a4paper,11pt]{article}
\pdfoutput=1 
\usepackage{jheppub} 
\usepackage[utf8]{inputenc}
\graphicspath{{./images/}}

\DeclareMathOperator{\Tr}{Tr}
\usepackage{slashed}

\newcommand{\lsim}{ \mathop{}_{\textstyle \sim}^{\textstyle <} }
\newcommand{\vev}[1]{ \left\langle {#1} \right\rangle }
\newcommand{\bra}[1]{ \langle {#1} | }
\newcommand{\ket}[1]{ | {#1} \rangle }

\def\diag{\mathop{\rm diag}\nolimits}
\def\SU{\mathop{\rm SU}}
\def\U{\mathrm{U}}
\def\tr{\mathop{\rm tr}}

\def\CA{{\cal A}}

\def\CG{{\cal G}}

\def\CL{{\cal L}}

\def\CM{{\cal M}}

\def\CO{{\cal O}}

\newcommand{\bC}{\mathbb{C}}
\newcommand{\bR}{\mathbb{R}}
\newcommand{\bZ}{\mathbb{Z}}

\def\SU{\mathrm{SU}}
\def\SO{\mathrm{SO}}
\def\tr{\mathop{\mathrm{tr}}\nolimits}
\def\Tr{\mathop{\mathrm{Tr}}\nolimits}

\def\vev#1{\langle#1\rangle}

\def\CP{\mathbb{CP}}

\def\bT{\mathbb{T}}

\def\beq#1\eeq{\begin{align}#1\end{align}}

\newcommand{\Kahler}{K\"ahler }

\newcommand{\modA}{\mathbb{E}}

\newcommand{\khalf}{k\to k+1}

\newcommand{\el}{\mathsf{e}}

\title{From 4d Yang-Mills to 2d $\CP^{N-1}$ model: \\ IR problem and confinement at weak coupling}

\preprint{IPMU17-0059}

\author{Masahito Yamazaki and Kazuya Yonekura}
\affiliation{Kavli IPMU (WPI), UTIAS, The University of Tokyo,  Kashiwa, Chiba 277-8583, Japan}

\abstract{
We study four-dimensional $\SU(N)$ Yang-Mills theory on $\bR \times \bT^3=\bR \times S^1_A \times S^1_B \times S^1_C$, with a twisted boundary condition by a $\mathbb{Z}_N$ center symmetry imposed on $S^1_B \times S^1_C$. This setup has no IR zero modes and hence is free from IR divergences which could spoil trans-series expansion for physical observables. Moreover, we show that the center symmetry is preserved at weak coupling regime. This is shown by first reducing the theory on $\mathbb{T}^2=S_A \times S_B$, to connect the model to the two-dimensional $\mathbb{CP}^{N-1}$-model. Then, we prove that the twisted boundary condition by the center symmetry for the Yang-Mills is reduced to the twisted boundary condition by the $\mathbb{Z}_N$ global symmetry of $\CP^{N-1}$. There are $N$ classical vacua, and fractional instantons connecting those $N$ vacua dynamically restore the center symmetry. We also point out the presence of singularities on the Borel plane which depend on the shape of the compactification manifold, and comment on its implications.
} 

\begin{document}

\maketitle
\section{Introduction}

\subsection{Confinement and resurgence}
It has been a long-standing problem to show that 
a non-trivial quantum Yang-Mills theory exists on the flat space $\mathbb{R}^4$, and that it
has a mass gap.
This is a challenging problem since Yang-Mills theory, while asymptotic free in the UV, is intrinsically strongly coupled in the IR.

One possible approach is to start with the perturbative expansion of the theory:
this is the best-understood part of the theory, can also be formulated 
mathematically \cite{Costello_Book}. One might then hope that a suitable re-summation of such a perturbative series
might lead to a complete theory \cite{tHooft:1977xjm}.

Of course, Yang-Mills theory on $\mathbb{R}^4$ is strongly-coupled in the IR and hence the perturbative expansion breaks down.
One might nevertheless try to avoid this problem by compactifying the theory onto e.g.\ a circle $S^1$ of radius $L$.
When  the radius is smaller than the dynamical scale $\Lambda$ of the theory ($L \ll \Lambda^{-1}$),
the renormalization-group-running of the gauge coupling constant stops at the scale $1/L$, 
at which the gauge coupling is small.
One then hopes to do a reliable weak coupling computation, perturbatively in the gauge coupling constant.
Once we have done this, one might hope to adiabatically continue back to the flat space.

The hope is that such an adiabatic continuation can be achieved by the theory of Borel-$\acute{\textrm{E}}$calle re-summation and resurgence \cite{Ecalle1,Ecalle2} (see e.g.\ \cite{Costin_Book,Sauzin_Book} for recent exposition).
This states that the perturbative series (say of the vacuum expectation value of an operator $\mathcal{O}$) should be thought of as part of 
the so-called trans-series expansion, containing both perturbative and 
non-perturbative corrections: 
\begin{align}
\langle \mathcal{O} \rangle
= \sum_{k=0}^{\infty} c_{0,k} g^k+ \sum_{I} e^{-\frac{S_I}{g^2}} \left( \sum_{k=0}^{\infty} c_{I,k} g^k \right)+ \dots \;,
\label{eq:trans_series}
\end{align}
where $I$ is a label for the saddle points of the action $S$, and $S_I$ denotes the value of the action at the $I$-th saddle point;
the coefficients $c_{I,k}$ represent the perturbative expansions around the $I$-th saddle point.
Assuming that we know all the saddle points contributing to the path integral,
all the coefficients $c_{0,k}$ and $c_{I,k}$ can be computed by perturbative methods,
and we obtain a trans-series as in \eqref{eq:trans_series}.
Now one hopes to turn the trans-series into a well-defined function 
$\langle \mathcal{O} \rangle(g)$ as a function 
of the coupling constant with the help of the resurgence theory,\footnote{The word resurgence in the physics literature sometimes refers to a stronger statement. Namely,  large-order asymptotic growth (as $k$ large) of the perturbative coefficients $c_{0,k}$ around the trivial saddle point contains the information of the non-perturbative saddle points (inside the same topological sector).
This has been discussed in the literature since long ago \cite{Bogomolny:1980ur,ZinnJustin:1981dx}.
See e.g., \cite{Pasquetti:2009jg,Aniceto:2014hoa,Honda:2016mvg,Honda:2016vmv,Dunne:2016jsr,Fujimori:2017oab} for various points of view on this issue.
}
and if that function is indeed resurgent (as might be expected from resurgence theory), i.e.\
endlessly continuable, we can adiabatically continue back to
the large value of the coupling constant along a suitable choice of path in the complex plane.

\paragraph{Challenges in resurgence}

There are problems with such an optimistic scenario.

First, such a compactification of the theory can dramatically change the physics; for example,
if we compactify the theory on a temporal circle we expect to encounter the confinement--deconfinement transition,
and thus the weak coupling computation is possible only in the deconfined phase.
One way to avoid such a phase transition is to ensure the existence of the center symmetry (more on this later in this paper).
This can be achieved, for example, 
by twisted boundary conditions or by suitable deformation of the Lagrangian.
Such a continuity from small to large size of the compactified direction is called adiabatic continuity. It was considered in physics independently 
and later combined with the mathematical idea of resurgence:
see e.g.,~\cite{
Unsal:2007vu,Kovtun:2007py,Unsal:2007jx,Unsal:2008ch,Shifman:2009tp,Shifman:2008ja,Argyres:2012ka,Argyres:2012vv,Dunne:2012ae,Dunne:2012zk,Poppitz:2012sw,Anber:2013doa,Cherman:2013yfa,Cherman:2014ofa,Dunne:2015ywa,Misumi:2014bsa,Misumi:2014jua, Dunne:2016nmc,Cherman:2016hcd,Sulejmanpasic:2016llc}.
(See also literature on large $N$ twisted Eguchi-Kawai model \cite{GonzalezArroyo:1982ub,GonzalezArroyo:1982hz,GonzalezArroyo:2010ss}).

Second, in order to cancel the ambiguity of Borel re-summation of perturbative series, we need to have a corresponding 
semi-classical saddle point for 
a singularity of the Borel plane on the positive real axis. 
In particular, it was argued that there is the renormalon pole \cite{tHooft:1977xjm}, whose action is of order $1/N$ 
of that of the instanton in the large $N$ limit.
Recently there has been huge progress in this respect (together with the adiabatic continuity), in the context of recent connection with resurgence. The activities in this field has been triggered by the works of \cite{Dunne:2012ae,Dunne:2012zk} on 2d $\mathbb{CP}^{N-1}$-model and \cite{Argyres:2012ka,Argyres:2012vv} for 4d adjoint QCD.\footnote{However, see \cite{Anber:2014sda} which argued that the usual renormalon diagrams
do not give any factorial growth of the coefficients of perturbative series in the setup of \cite{Argyres:2012ka,Argyres:2012vv}. } 


While these points are of relevance to this paper, our main focus here is yet another, although related,
source of trouble for the resurgence program, associated with IR divergences. The problem of IR divergences is already
stressed in the context of resurgence in \cite{Dunne:2012ae}, but we discuss this problem again by emphasizing 
the point that they spoil the formal trans-series expansion of physical observables.

\subsection{The problem of the infrared divergences}\label{subsec:IR}

In this paper we point out and study the subtlety which spoils the resurgence program: the issue of IR divergence.
In short, the problem is that the perturbative (or more generally trans-series) expansion \eqref{eq:trans_series} in itself is ill-defined 
if the observable suffers from IR divergences.

\paragraph{Instanton computation of vacuum energy.}
To illustrate the issue of IR divergence, let us study the theta-angle dependence of the vacuum energy of 
the Yang-Mills theory on $\mathbb{R}^4$.
Instanton computation gives the formula for the $\theta$-dependence of vacuum energy $E(\theta)$ as \cite{tHooft:1976snw}
\beq
E(\theta) \sim -\int^\infty_0 \frac{d \rho}{\rho^5} (\mu \rho)^{b_1} e^{-\frac{8\pi^2}{g^2(\mu)}} \cos \theta \;,
\label{E_instanton}
\eeq
where $\mu$ is an arbitrary renormalization scale, $b_1$ is the coefficient of the one-loop beta function, $\rho$ is the size modulus of the instanton, and $g(\mu)$ is the running gauge coupling constant at the scale $\mu$.
This integral is divergent in the IR region $\rho \to \infty$. 
Still, one might hope that the formula gives qualitatively right answer
by introducing naive IR cutoff, which we may take to be the dynamical scale $\Lambda$ of the theory.
This gives the answer\footnote{The minus sign is chosen for consistency with the result of Vafa and Witten~\cite{Vafa:1984xg} that the 
lowest energy is realized at $\theta=0$.}
\begin{align}
E(\theta) \sim - \Lambda^4 \cos \theta \;.
\label{eq:E_weak}
\end{align}
It turns out, however, that this is not the correct result. Witten argued~\cite{Witten:1980sp,Witten:1998uka} (see the first section of \cite{Witten:1998uka} for a beautiful summary; see also \cite{Giusti:2007tu,Unsal:2012zj,Yonekura:2014oja} for further support) 
that in the large $N$ limit, there are infinitely-many
metastable vacua labelled by an integer $\el \in \bZ$, and the vacuum energy of each of the metastable vacua are given by
\beq
E_\el(\theta) \to  \Lambda^4 (\theta -2\pi \el)^2~~~~~(N \to \infty) \;.
\eeq
Note that each $E_\el(\theta)$ is not even a periodic function of $\theta$! 
We can recover the periodicity by considering the energy of the true vacuum, namely minimum of all the $E_{\el}$'s:
\begin{align}
E(\theta)= \min_{\el} E_\el(\theta) \;,
\end{align}
however this is discontinuous at $\theta=\pm \pi$. 
(See \cite{Gaiotto:2017yup} and references therein for the physics at $\theta = \pm \pi$.) 
The actual situation is that each vacuum labeled by $\el \in \bZ$ is 
metastable (in the large $N$ limit),
and the true stable vacuum $\el=0$ at $\theta=0$ is adiabatically continued to a metastable vacuum under the monodromy $\theta \to \theta+2\pi$.
This is very different from the naive answers expected from the instanton calculus \eqref{E_instanton}, even at the qualitative level.

\paragraph{Linde's problem.}
Let us next consider the thermal free energy $F$ defined by
\beq
\exp(-F \cdot {\rm Vol.}) = \Tr e^{- \beta H} = \int_{S^1_\beta \times \bR^3} [DA]\, e^{-S} \;,
\eeq
where $\beta=T^{-1}$ is the inverse temperature and ${\rm Vol.}$ is the space-time volume,
which we regularize to be a finite number by introducing some IR cutoff.
After the compactification on the thermal circle $S^1_\beta$, the 4d Yang-Mills is reduced to 3d Yang-Mills up to Kaluza-Klein (KK) modes
with the 3d gauge coupling given by $g_3^2=Tg^2$. The contribution of 3d Yang-Mills to the free energy is expanded by $g_3^2=Tg^2$,
but it has a mass dimension. Thus the expansion is of the form
\beq
\beta F \sim \text{[KK contribution]}+  \sum_{n =0}^\infty a_n  \frac{(Tg^2)^{n+3}}{(m_{\rm IR})^n}\;.
\eeq
where $\text{[KK contribution]}$ means the contributions from nonzero modes on $S^1_\beta$, and the expansion by $Tg^2$ comes from 3d Yang-Mills.
In naive perturbation theory, $m_{\rm IR}=0$ and the above expansion has severe infrared problem.
In improved perturbation theory, $m_{\rm IR}$ may be taken to be the thermal mass of the fields. However, for the 3d gauge fields,
the thermal mass is known to be at most $m_{\rm thermal} \lsim Tg^2$. Therefore, in the above expansion, higher loops do not give higher powers of 
the coupling $g$ and hence all the higher loops give the fixed order  $\CO(g^6)$ of the coupling expansion (in the most optimistic case
of $m_{\rm thermal} \sim Tg^2$). The perturbative expansion, and hence the (trans)series expansion, is therefore  not well-defined \cite{Linde:1980ts}.

\bigskip
The naive cutoff prescription of the IR divergence does not work 
in each of the two cases discussed above.
Of course, some observables (IR safe observables) do not depend on the details of IR regularizations,
but many observables of physical interest are sensitive to such subtleties.

We stress that the IR problems mentioned above are conceptually different from (although related to) the problem of 
whether the coupling $g$ is large or small and we can numerically trust the weak coupling results or not.\footnote{For example, even for weakly coupled QED, there are IR divergences due to soft photons. Therefore the problem of IR divergences are conceptually different from the strong dynamics. However, in the case of soft photons, physical treatment of IR divergences are not so difficult.} The issue is whether or not the coefficient in the perturbative (or trans-series) expansion is well-defined.
Even if $g$ is very large, the resurgence might give a sensible answer as long as the trans-series expansion is well-defined.
But if there is IR divergences, there is no way to do resurgence from the beginning.

The question is then to find a concrete setup in which the trans-series expansion is well-defined. 
Doing this for the pure $\SU(N)$ Yang-Mills theory\footnote{One can also add adjoint fermions, and also fundamental fermions whose flavor number is
a multiple of $N$.} is the motivation for the next section.

\section{Our setup}

\begin{figure}[htbp]
\centering\includegraphics[scale=0.15]{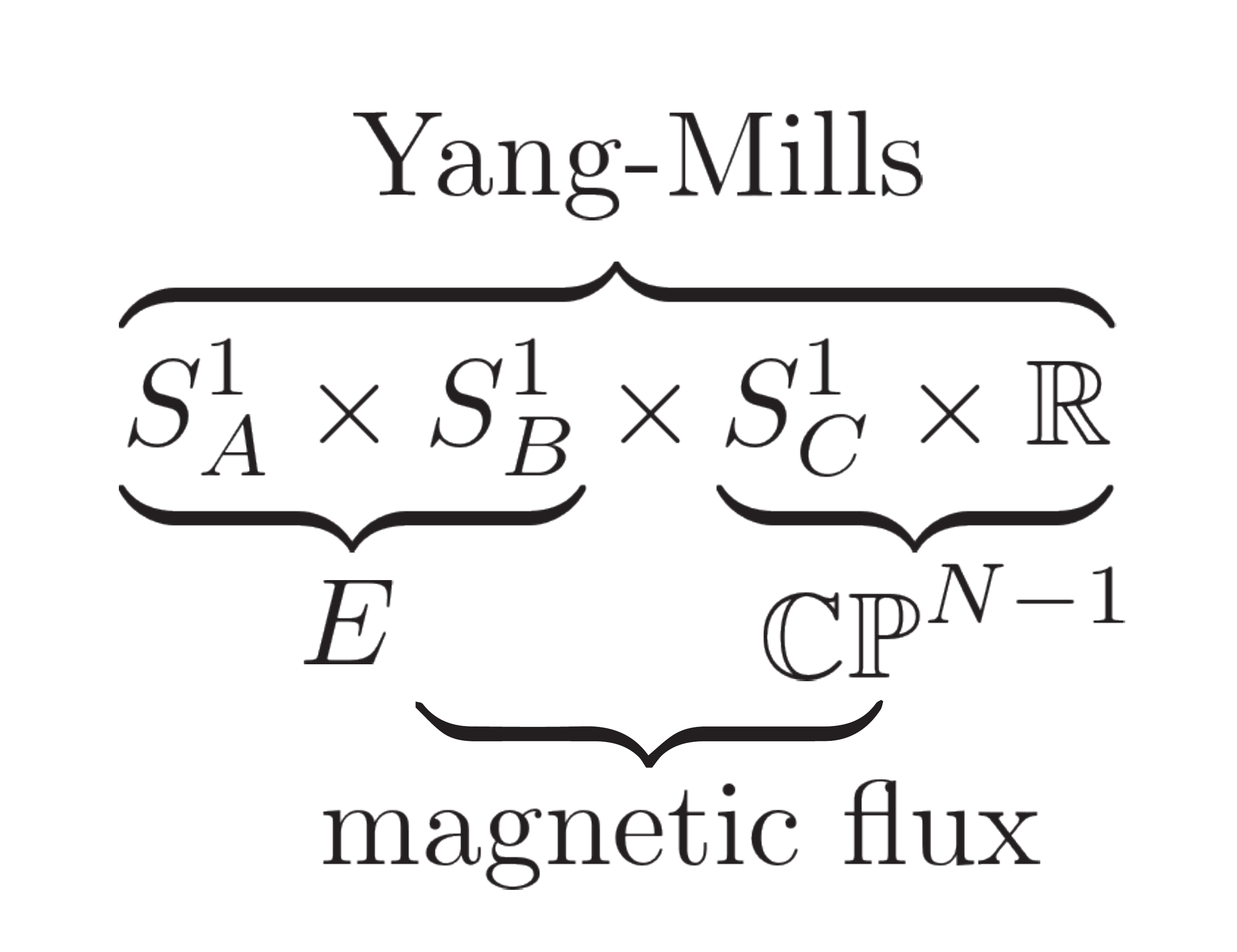}
\caption{We consider Yang-Mills theory on the geometry $\bR \times S^1_A\times S^1_B\times S^1_C$.
When the torus $E=S^1_A\times S^1_B$ is small, we obtain the two-dimensional $\mathbb{CP}^{N-1}$-model
in the remaining directions $\mathbb{R}\times S^1_C$. We include a unit 't Hooft magnetic flux (i.e., twisting by the center symmetry) along $S^1_B\times S^1_C$,
which plays rather crucial roles in this paper.}
\label{fig:circles}
\end{figure}

In this paper we study the pure Yang-Mills theory with gauge group $\SU(N)$,
with gauge coupling constant $g$ and theta angle $\theta$. Our setup is summarized in Figure \ref{fig:circles}.

\subsection{Yang-Mills theory in a box}

In this paper the pure Yang-Mills theory is put into a box, namely compactified on 
a spatial three-torus 
\begin{align}
\mathbb{R}\times \mathbb{T}^3=\mathbb{R}\times  S^1_A\times S^1_B \times S^1_C \;,
\end{align}
where the non-compact direction $\mathbb{R}$ is the temporal direction.
Here $S^1_{A,B,C}$ are spatial directions.
We will denote their circumferences of three spatial directions by $L_{A,B,C}$, respectively.
These circumferences play the role of the IR cutoff, and the limit to the flat space $\mathbb{R}^4$ is 
achieved by $L_{A,B,C}\to\infty$ with their ratios kept finite.

In this paper we will primarily study the parameter region
\begin{align}
 L_{A}, L_{B} \ll L_C \;,
\label{con2}
\end{align}
with the relative size of $L_A$ and $L_B$ kept finite.
Note this breaks the symmetries between $S^1_{A,B}$ and $S^1_C$.

The hierarchy in scales \eqref{con2} means that below energy scales $1/L_A$ and $1/L_B$ the theory is effectively described by 
the two-dimensional theory on $\mathbb{R}\times S^1_C$.
We show that the resulting two-dimensional theory is given by the two-dimensional sigma model
whose target space is $\mathbb{CP}^{N-1}$. This connection with the two-dimensional physics 
will be rather useful for understanding the physics of four-dimensional Yang-Mills theory.

Notice that the parameter region \eqref{con2} is different from the decompactification limit of $\mathbb{R}\times \mathbb{T}^3$  into $\mathbb{R}^3\times S^1$,
where one might encounter a version of Linde's problem:
\begin{align}
L_{A}, L_{B} \gg L_C \;.
\label{con1}
\end{align}
However, one can hope that after a suitable resurgence analysis in the region  \eqref{con2}
we may extrapolate the answer to the region \eqref{con1}. 
We will give some comments on this point in section~\ref{sec:BorelLinde}.

\subsection{\texorpdfstring{Twist by 1-form $\bZ_N$ symmetry}{Twist by 1-form ZN symmetry}}\label{sec:oneform}

As another crucial ingredient for our construction, we consider twisted boundary condition along $S^1_B \times S^1_C$.
The four-dimensional pure Yang-Mills theory has a center symmetry, or in modern language a global ``electric'' one-form $\mathbb{Z}_N$-symmetry~\cite{Gaiotto:2014kfa}. In the Language of \cite{Gaiotto:2014kfa}, 
the one-form symmetry twist along $S^1_B \times S^1_C$ used in this paper is generated by the 
surface operator supported on $\text{(pt)} \times S^1_A\times \mathbb{R}$, where $\text{(pt)} \in S^1_B \times S^1_C$ is a point.
(See appendix \ref{app:one_form} for quick summary of one-form symmetry).
As we will see later, this global symmetry acts on the Wilson lines along the cycle $S^1_B$. 

This twist is crucial for the following reason. First, we will see that this ensures that 
\begin{itemize}\parskip=-2pt
\item gauge symmetry is completely broken at classical vacua around which we perform perturbation, hence there is no 
massless modes in the IR.
\end{itemize}
We therefore do not encounter a subtle issue of IR divergences.
Second, we will see later in this paper (section \ref{sec:center_restore}) that while the one-form center symmetry is broken classically,
it is restored after quantum tunneling effects are taken into account:
\begin{itemize}\parskip=-2pt
\item center symmetry is preserved.
\end{itemize}
Since the presence of the one-form center symmetry is proposed as a characterization of the 
confinement phase~\cite{Gaiotto:2014kfa}, we conclude that our system confines,
at least in the parameter regions \eqref{con2}. We expect that this is true for all $L_{A},L_{B},$ and $L_{C}$ 
as long as their ratios are kept finite. (The case of taking $L_{A,B}/L_C \to \infty$ is commented in section~\ref{sec:BorelLinde}.)

Such a dynamical restoration of center symmetry is shown by reducing the theory onto the $\mathbb{R}\times S^1_C$-directions, 
and matching the resulting theory with the two-dimensional $\mathbb{CP}^{N-1}$-model with 
 $\mathbb{Z}_N$-twisted boundary condition studied recently in connection with resurgence~\cite{Dunne:2012ae,Dunne:2012zk,Sulejmanpasic:2016llc}. Namely, we show that the following two twists are the same after the reduction from four to two dimensions:
\begin{itemize}
\item the twist in the $\SU(N)$ Yang-Mills theory by the one-form symmetry on $S^1_B \times S^1_C$,
\item the twist in the $\CP^{N-1}$ model by the $\bZ_N$ global symmetry on $S^1_C$.
\end{itemize}
For this reason we will spend the next two sections exploring the connection between four-dimensional Yang-Mills theory
and the two-dimensional $\mathbb{CP}^{N-1}$-model.

\section{\texorpdfstring{From $\SU(N)$ Yang-Mills to $\CP^{N-1}$ }{From SU(N) Yang-Mills to CP(N-1)}}

Let us first consider the compactification of the pure $\SU(N)$ Yang-Mills theory on the two-dimensional torus (elliptic curve) $E=\bT^2=S^1_A\times S^1_B$ down to two dimensions.

When the Yang-Mills theory is compactified on $E$ and the Kaluza-Klein modes are integrated out,
the light degrees of freedom on $E$ are the zero-energy configuration of gauge fields, {\it i.e.} 
flat connections on $E$,  except near singular points of the moduli space of flat connections
where there is an enhanced symmetry and W-bosons become light. 
The moduli space of flat $\SU(N)$ connections on $E$ is known to be
the projective space 
$\CP^{N-1}$ at the holomorphic level~\cite{Looijenga1,Looijenga2,Friedman:1997yq}. 
Here we review and spell out some details of this relation between flat connections and $\CP^{N-1}$.

\subsection{Yang-Mills flat connections on a torus}
We choose a complex coordinate of the torus $E=\bT^2$ to be 
\beq
z \sim  z+1 \sim z+\tau \;,
\eeq
where $\tau $ is a complex parameter (modulus of torus) with ${\rm Im}(\tau)>0$. The torus $E$ has the flat metric 
\beq
ds^2 = L^2_E |dz|^2 \;,
\eeq
where $L_E$ is the size of the torus. In the notation of the previous section, 
these parameters are given as $\tau=i (L_B/L_A)$ and $L_E=L_A$. But it is not difficult to take $\tau$ to be more general complex numbers.

We call the cycles on $E$ corresponding to $z \to z+1$ and $z \to z+\tau$ as
the A-cycle and B-cycle, respectively. In our previous notation these are $S^1_A$ and $S^1_B$, respectively.

The gauge field on $E$ can be represented as $ A=A_{\bar{z}}d\bar{z}+A_z dz$
in the complex coordinate $z$.
We take $A$ to be anti-hermitian so that the field strength is given by $F=dA+A\wedge A$.
For flat connections, the $\bar{z}$-component $A_{\bar{z}}$ is expanded as
\beq
A_{\bar{z}} = - (A_z)^\dagger \xrightarrow{\text{flat connection}} \frac{-2\pi i}{ (\tau - \tau^*)}\diag(\phi_1, \cdots, \phi_N) \;,\label{eq:flat}
\eeq
where $\phi_i$ are complex scalars which are independent of $z$ and satisfy $\sum_{i=1}^N \phi_i=0$.
The constant prefactor $-2\pi i/(\tau-\tau^*)$ is chosen for later convenience. 
Wilson lines in the A- and B-cycles are given by
\beq
\begin{split}
& U_A =\exp  \int_A (-A_{\bar{z}}d\bar{z} - A_z dz)=\diag\left( \cdots, \, \exp\left(2\pi i \cdot \frac{\phi_i -\phi_i^*}{\tau - \tau^*} \right) , \cdots \right)  ,  \\
& U_B =\exp  \int_B (-A_{\bar{z}}d\bar{z} - A_z dz)=\diag\left( \cdots, \, \exp\left(2\pi i \cdot \frac{\tau^*\phi_i -\tau\phi_i^*}{\tau - \tau^*} \right) , \cdots \right).\label{eq:Wloop}
\end{split}
\eeq
The scalars $\phi_i$ have periodicities
\beq
\phi_i \sim \phi_i +(\tau m_i - n_i) \;,~~~m_i,n_i \in \bZ \;,~~~\sum_{i}m_i = \sum_{i}n_i =0 \;,
\eeq
and thus can be regarded as a point on $E$ with the constraint $\sum_i \phi_i=0$.
Furthermore, $\phi_i$'s are permuted by the Weyl group of $\SU(N)$,
namely the $N$-th symmetric group ${\mathfrak S}_N$:
an element of the Weyl group $\sigma \in {\mathfrak S}_N$ acts as $\phi_i \to \phi_{\sigma(i)}$.
Therefore, the moduli space of flat connections is given by
\beq
& \CM_{\rm flat} =\modA / {\mathfrak S}_N, \\
& \modA := \left\{ (\phi_1,\cdots, \phi_N) \in E^N; \sum_i \phi_i=0 \right\} .\label{eq:abel}
\eeq

\begin{figure}[htbp]
\centering\includegraphics[scale=0.35]{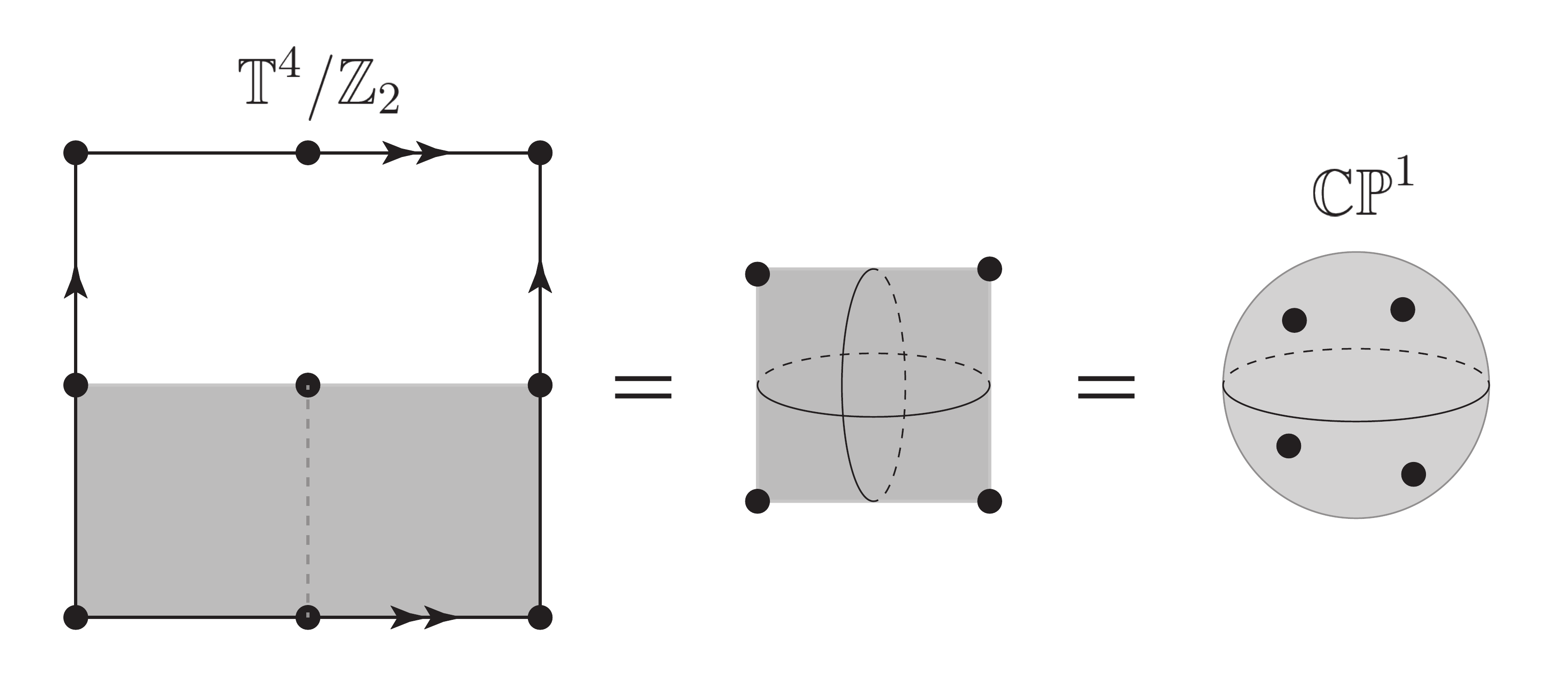}
\caption{For the case $N=2$, the moduli space of flat connections $ \CM_{\rm flat}$ on $E=\mathbb{T}^2$ is $\mathbb{T}^2/\mathbb{Z}_2$, which as in this figure can be 
identified with $\mathbb{CP}^1$. The four fixed points of $\mathbb{T}^4/\mathbb{Z}_2$ are mapped into the four singular points on $\mathbb{CP}^1$, represented by black dots.}
\label{fig:T4Z2}
\end{figure}

The case of $N=2$ is particularly simple, since then we have 
\beq
& \modA := \left\{ (\phi_1,-\phi_1) \in E^2 \right\} =E\: , \quad \CM_{\rm flat} =\mathbb{T}^2/\mathbb{Z}_2 \;.
\eeq
One can then see as in Figure \ref{fig:T4Z2} that this is indeed $\mathbb{CP}^1=S^2$
as an algebraic variety. What has to be kept in mind is that the manifold has four singular points, 
where extra massless W-bosons appear.

The case of $N>2$ is more complicated, but
there is a convenient way to describe such $N$ points $\phi_i$ on $E$. Consider a space of meromorphic functions $F$ which
are allowed to have $N$-th order pole at $z=0$. Namely, $F$ is allowed to behave as 
$F(z) \sim  z^{-N}$ ($z \sim 0$). We denote such a space of meromorphic functions on $E$
as $H^0(E,\CO(Np))$, where $p$ represents the point $p=\{ z=0 \in E \}$, and $\CO(Np)$ indicates that
the functions are allowed to have $N$-th order pole at $p$.\footnote{
More precisely, $\CO(Np)$ is a line bundle on $E$ associated to the divisor $Np$. See any textbook
on algebraic geometry (e.g.\ \cite{GH}) for the meaning of those words ``line bundle" and ``divisor". The reader who is not familiar with algebraic geometry
can neglect them.} 
When $F \in H^0(E,\CO(Np))$, the Cauchy theorem for $dF/F=\partial_z F/F dz$ on $E$
implies that there must be $N$-points $q_i~( i=1,\cdots,N)$ 
at which $F(z)$ has a zero. This is because by the Cauchy theorem, the sum of residues of $dF/F$ must be zero
and it has the residue $+1$ at a simple zero of $F$ and $-1$ at a simple pole, with obvious generalization when zero or pole have multiplicity. 
If $q_i=p$ for some $i$, that means that the degree of the pole at $p=\{z=0\}$
is reduced by the number of $q_i$ for which $q_i=p$. If $q_i=q_j$, that means that the function $F$ has double (or more generally multiple) 
zero at that point. 

Then the claim is that the set of unordered points $\{\phi_i\}_{1 \leq i \leq N}$ 
on the torus $E$, that is $\CM_{\rm flat}=\modA/{\mathfrak S}_N$, is given by
${\mathbb P} H^0(E, \CO(N p)) $, where ${\mathbb P} H^0(E, \CO(N p))$ is the projective space
associated to the vector space $H^0(E, \CO(N p))$. The correspondence is given by
mapping the $N$ points $q_i$ of zeros of a function $F \in H^0(E, \CO(N p))$ to the points on $\CM_{\rm flat}=\modA/{\mathfrak S}_N$.
Also, the vector space $H^0(E, \CO(N p))$ has complex dimension $N$ and hence $H^0(E, \CO(N p)) \cong \bC^N$,
so we get $\CM_{\rm flat} \cong \CP^{N-1}$.
This can be shown by well-known techniques in algebraic geometry (see e.g.~\cite{Friedman:1997yq}).
Here we would like to give more or less elementary explanation.

\begin{figure}[htbp]
\centering\includegraphics[scale=0.45]{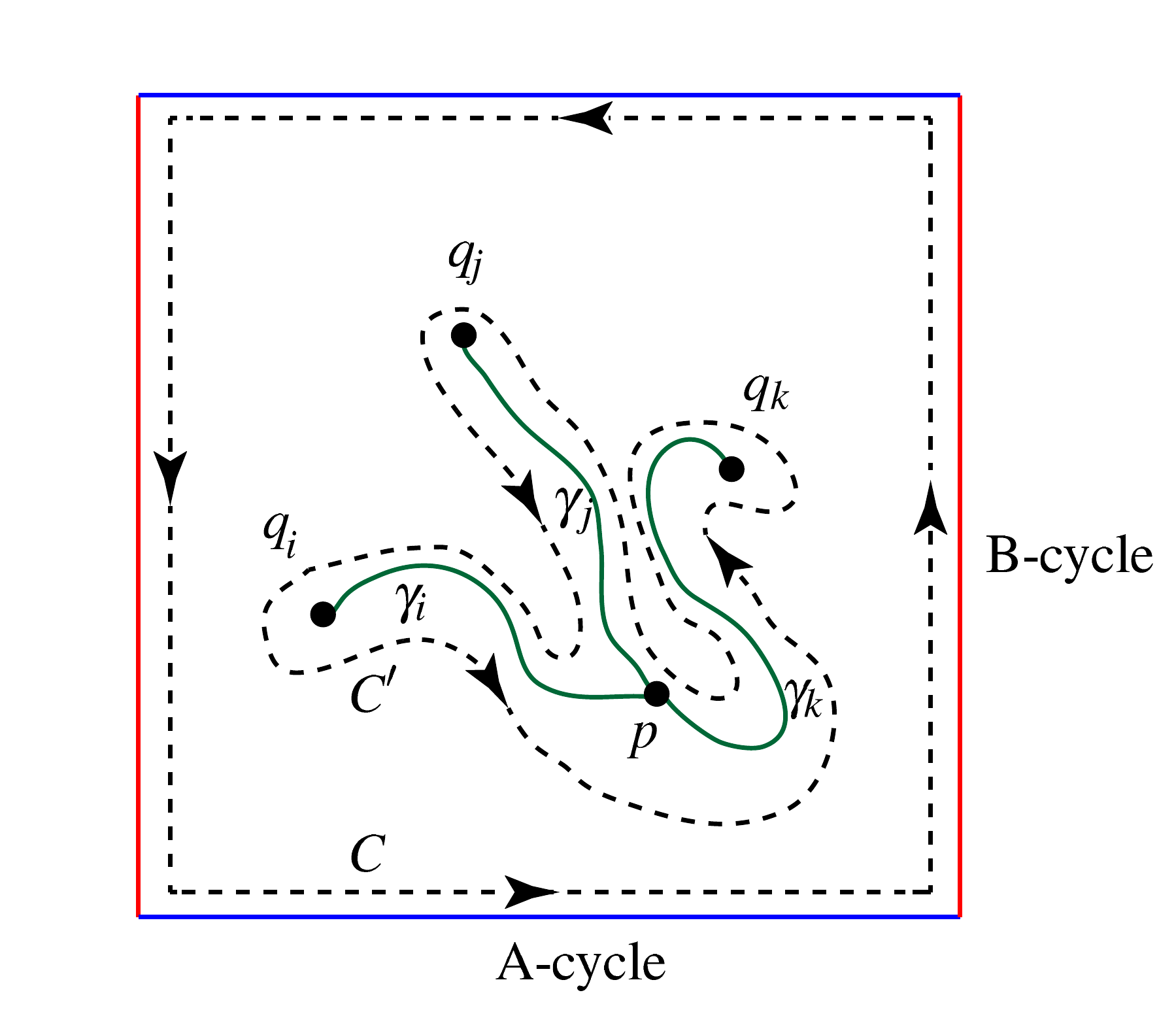}
\caption{The path $\gamma_i$ connects $p$ and $q_i$. Since this is a two-torus, the blue (red) edges are identified, 
and are the A-cycle and the B-cycle, respectively.
The contour $C=ABA^{-1}B^{-1}$ can be deformed into the contour $C'$, which can be decomposed into small cycles surrounding $\gamma_i$'s.}\label{fig:path}
\end{figure}

Let $\phi_i$ be the coordinates corresponding to $q_i$ associated to the zero points of a function $F$. 
Then, first we want to show that they satisfy $\sum_i \phi_i=0$ so that they define a point on $ \modA $ defined in \eqref{eq:abel}. One way to see
this is as follows. Corresponding to each $q_i$, take a path $\gamma_i$ from $p$ to $q_i$ such that $\gamma_i$ and $\gamma_j$
do not intersect other than at the point $p$. See Figure~\ref{fig:path}.
By eliminating $\gamma_i$ from $E$, the $\log F$ becomes well defined which has branch cuts at $\gamma_i$.
Consider the integration of $\log F dz$ on a loop of the form $ABA^{-1}B^{-1}$, where $A$ and $B$ are the A-cycle and B-cycle of the torus $E$.
Then we get a value of the form $2\pi i (\bZ +\tau \bZ) $.
This is because the contributions from $A$ and $A^{-1}$ add up to give $2\pi i n \int_A dz$ where $2\pi i n \in 2\pi i \bZ$ 
comes from the difference of $\log F$ before and after going around the loop $B$. 
Similarly, the contributions from $B$ and $B^{-1}$ add to give $2\pi i m \int_B dz$.
On the hand, applying the Cauchy theorem to this integral, we get $-\sum_i 2\pi i \int_{\gamma_i} dz$.
By using $\int_{\gamma_i} dz = (\phi_i - 0) $, we get the desired result $(\sum_i \phi_i ) \in (\bZ +\tau \bZ) $. 

From the results obtained above, 
we see that a meromorphic function $F \in H^0(E,\CO(Np))$ gives a data $\{\phi_i \}/ {\mathfrak S}_N$ 
needed for a flat connection.
Suppose that two meromorphic functions $F$ and $G$ have the same set of zero points $\{ \phi_i \}$.
Then, their ratio $F/G$ does not have either a zero or pole. Such a holomorphic function must be a constant, $F/G =\text{const}$.
This means the following. First, notice that the space of functions $H^0(E, \CO(N p))$ is a complex vector space.
Let ${\mathbb P} H^0(E, \CO(N p))$ be the projective space associated to this vector space. Namely, we introduce equivalence relation $F \sim cF$ 
for constants $c \in \bC - \{0\}$, and divide the space $ H^0(E, \CO(N p)) - \{ 0 \}$ by this equivalence relation.
Then, what we found above is that there is an injective map ${\mathbb P} H^0(E, \CO(N p)) \to \CM_{\rm flat}$.

It turns out that the map above is also surjective. Indeed, given a set of points $\{ \phi_i \}$ satisfying $\sum_i \phi_i=0$, we can explicitly construct a function $F \in H^0(E, \CO(N p))$
which has zeros at these points and an $N$-th order pole at $p=\{z=0\}$:
\beq
F(z) &= \prod_{m,n \in \bZ} f(z+m+n\tau)\;, \quad
f(z) =  \prod_{i=1}^N \frac{(z-\phi_i)}{z}\;.
\eeq
The infinite product is convergent if $\sum_i \phi_i=0$, because $\log f(z) \to  - z^{-1}\sum_i \phi_i + \CO(z^{-2})$ for $z \to \infty$,
and $\sum_{(m,n) } 1/(z+m+n\tau)^k$ is convergent for $k \geq 2$.

This establish that the moduli space of flat connections is given by
\beq
\CM_{\rm flat} = {\mathbb P} H^0(E, \CO(N p)) = \CP^{N-1}\;,
\label{eq:M_CPN}
\eeq
where we have used the fact that $H^0(E, \CO(N p))$ is an $N$-dimensional vector space, as can be seen by counting the number of
parameters to specify $F \in H^0(E, \CO(N p))$ (i.e., $\phi_i$ with $\sum_i \phi_i=0$ gives $N-1$ parameters and the overall scale gives one parameter.)

We remark that the equivalence \eqref{eq:M_CPN} should be interpreted as equivalence as algebraic varieties,
and at the classical level, the metric of the moduli space $\CM_{\rm flat}$ is not the Fubini-Study metric. 
However, this point does not affect our qualitative discussions in this paper.

\subsection{\texorpdfstring{Explicit map from $ \CM_{\rm flat}$ to $\CP^{N-1}$}{Explicit map from M(flat) to CP(N-1)}}\label{sec:explicit}
The above discussion was rather abstract. We now make the map $\CM_{\rm flat}=\modA / {\mathfrak S}_N \to \CP^{N-1}$ more explicit.

We denote the root lattice of 
$\SU(N)$ by ${\mathbb L}$
\beq
{\mathbb L}=\left\{ \vec{\ell} =(\ell_1, \cdots, \ell_N) \in \bZ^N; \sum_i \ell_i=0 \right\}\;. \label{eq:root} 
\eeq
The weight lattice is spanned by the fundamental weights
\beq
\vec{e}_k=(\overset{1}{1},\cdots,\overset{k}{1},0,\cdots,0) - \frac{k}{N}(1,\cdots,1)\;. \label{eq:weight}
\eeq

We define theta functions as
\beq
\theta_k(\vec{\phi}) :=  \sum_{ \vec{\ell} \in {\mathbb L}} e^{ \pi i \tau (\vec{\ell}+\vec{e}_k)^2+ 2\pi i (\vec{\ell}+\vec{e}_k) \cdot \vec{\phi}  } ~~~~~(k=1,\cdots,N)\;,
\label{eq:Atheta}
\eeq
where $\vec{\phi}=(\phi_1,\cdots,\phi_N)$ and the inner product between vectors is defined as $\vec{\phi}\cdot \vec{\ell}=\sum_i \phi_i \ell_i$. 
These theta functions are invariant under the Weyl symmetry ${\mathfrak S}_N$ acting on $\vec{\phi}$, because each set
\beq
\vec{e}_k+{\mathbb L}=\{ \vec{e}_k+\vec{\ell} ; \vec{\ell} \in {\mathbb L} \}
\eeq
is Weyl invariant, and the Weyl symmetry preserves the inner product. Furthermore, under the shift 
\beq
\vec{\phi} \to \vec{\phi} +\tau \vec{m} - \vec{n}~~~~~(\vec{m}, \vec{n} \in {\mathbb L}) \;,
\eeq
they transform as
\beq
\theta_k(\vec{\phi} + \tau \vec{m} - \vec{n}) =e^{ -\pi i \tau \vec{m}^2-2\pi i \vec{m} \cdot \vec{\phi}}\theta_k(\vec{\phi}) \;.\label{eq:thetatransl}
\eeq
Note that the factor $e^{ -\pi i \tau \vec{m}^2-2\pi i \vec{m} \cdot \vec{\phi}}$ is independent of $k$.

We denote points of $\CP^{N-1}$ by using homogeneous coordinates as $[Z_1,\cdots,Z_N]$.
Then, if we define 
\beq
\varphi (\vec{\phi}):=[\theta_1(\vec{\phi}), \cdots, \theta_N(\vec{\phi})] \;,
\eeq
then the above properties imply that this is a well-defined map from $\CM_{\rm flat}$ to $\CP^{N-1}$.\footnote{The functions $\theta_k~(k=1,\cdots,N)$ are expected to
be not simultaneously zero at any value of $\vec{\phi}$, since there are $N$ functions of $N-1$ variables. For more 
mathematical justifications, see the paragraph below.}
We claim that this is an isomorphism between $\CM_{\rm flat}$ and $\CP^{N-1}$.

In the above discussion, we have given $\theta_k$ explicitly. 
In fact, there is an algebraic geometric reason for the existence of such a map. 
(The reader who is not interested in algebraic geometry can skip this paragraph.) Let $\modA$ 
be the space of flat connections before dividing by the Weyl group as defined in \eqref{eq:abel}.
In the previous subsection, we established that $\modA/ {\mathfrak S}_N \cong \CP^{N-1}$. Thus there is a natural projection map
\beq
\pi : \modA \to \CP^{N-1}.
\eeq
On the $\CP^{N-1}$, there is a line bundle $\CO(1)$ (which is also written as $\CO(H)$ for a hyperplane $H$ in $\CP^{N-1}$
and is called the hyperplane line bundle).
It has the properties that its holomorphic sections are one-to-one correspondence with linear functions of the homogeneous coordinates 
$Z_k~(k=1,\cdots,N)$ of the $\CP^{N-1}$.
With this identification, there are $N$ linearly independent sections $Z_k$ of $\CO(1)$. Now, we can pull-back this line bundle
by $\pi$ to define a line bundle $L=\pi^* \CO(1)$ on $\modA$. Then we also get $N$ linearly independent sections $\theta_k=\pi^* Z_k$ of
the line bundle $L$ by the pull-back. By using these sections, we can define a map 
\beq
\tilde{\varphi}:\modA \ni \vec{\phi} \mapsto [\pi^* Z_1(\vec{\phi} ),\cdots,\pi^* Z_N (\vec{\phi} ) ] \in \CP^{N-1}.
\eeq
By construction, this map coincides with the original projection map $\pi$. 
On the other hand, $\modA$ is an Abelian variety because it is equivalent to $E^{N-1}$ as is clear from \eqref{eq:abel}. 
Also, $L$ is a positive line bundle in the sense that its curvature 
can be positive definite by choosing an appropriate metric on $L$. 
There is a general classification and explicit construction of such line bundles and their sections on Abelian varieties.
See the section on ``Theta Functions" in~\cite{GH}. These are precisely the theta functions $\theta_k(\vec{\phi})$ given in \eqref{eq:Atheta}.
In fact, the nontrivial transformation \eqref{eq:thetatransl} implies that $\theta_k$ are not functions, but sections of a line bundle $L$.

\subsection{\texorpdfstring{1-form $\bZ_N$ symmetry in $\mathbb{CP}^{N-1}$}{1-form Z(N) symmetry in CP(N-1)}}\label{sec:ZN_CPN}

By compactification of space-time on $S^1$, a 1-form symmetry splits to a 1-form symmetry and a 0-form symmetry.
Here, ``0-form" symmetry means a usual global symmetry.
So, we get a usual global symmetry for each $S^1$. 

In the context of reducing four dimensional Yang-Mills to two dimensional $\CP^{N-1}$, 
we are interested in the two 0-form symmetries
$\bZ^{(A)}_N$ and $\bZ^{(B)}_N$ associated with the torus $E=\bT^2=S^1_A\times S^1_B$.
Their action on the Wilson lines $U_A$, $U_B$ along $S^1_A$, $S^1_B$ is described as
\beq
(p,q)  \in \bZ^{(A)}_N \times \bZ^{(B)}_N : (U_A, U_B) \mapsto \left(e^{-\frac{2\pi i p}{N}} U_A, e^{-\frac{2\pi i q}{N}} U_B \right) ,\label{eq:YM1form}
\eeq
where the minus signs are introduced just for later convenience. Notice that the two $\bZ_N$'s, $\bZ^{(A)}_N$ and $\bZ^{(B)}_N$ commute with each other.

Let us translate the action of $\mathbb{Z}_N$-symmetries $\bZ_N^{(A)}$ and $\bZ_N^{(B)}$
in \eqref{eq:YM1form} into the language of $\mathbb{CP}^{N-1}$-model.

By using the relation \eqref{eq:Wloop}, we find the $\mathbb{Z}_N$-symmetries
in \eqref{eq:YM1form} act on $\phi_k$ as
\beq
(p,q)  \in \bZ^{(A)}_N \times \bZ^{(B)}_N : \phi_k \mapsto \phi_k -\frac{p\tau - q}{N}+  \tau \bZ +\bZ \;.
\eeq
where $ \tau \bZ +\bZ$ means an arbitrary constant of the form $\tau m_i -n_i$ such that $\sum_i \phi_i=0$ is preserved.
By defining a vector 
\beq
\vec{c}_j=(0,\cdots,0,\overset{j}{1},0,\cdots,0) - \frac{1}{N}(1,\cdots,1)=\vec{e}_{j}-\vec{e}_{j-1} \;. \label{eq:centervector}
\eeq
the above action is written as
\beq
\vec{\phi} \mapsto \vec{\phi}+(p\tau - q) \vec{c}_j
\eeq 
for arbitrary $j$. The reason that $j$ can be arbitrary is because $\vec{c}_j - \vec{c}_{j'} \in {\mathbb L}$ for any $j$ and $j'$.

Now we can study the action of them on the theta functions $\theta_k$ defined in Sec.~\ref{sec:explicit}.
First, one can check that for $(p,q)=(0,1)$, the change of the theta functions is given as
\beq
\theta_k( \vec{\phi} -  \vec{c}_j) = e^{\frac{2\pi i k}{N}}\theta_k( \vec{\phi}) \;.
\eeq
Next, for $(p,q)=(1,0)$, we get
\beq
\theta_k( \vec{\phi}+  \tau \vec{c}_j)=
 \sum_{ \vec{\ell} \in {\mathbb L}} 
 e^{ \pi i \tau (\vec{\ell}+\vec{e}_k+\vec{c}_j)^2+ 2\pi i (\vec{\ell}+\vec{e}_k+\vec{c}_j) \cdot \vec{\phi} - \pi i \tau \vec{c}_j^2 -2\pi i \vec{c}_j \cdot \vec{\phi}  } \;.
\eeq
Now, note that $\vec{e}_k+\vec{c}_{k+1}=\vec{e}_{k+1}$ and $\vec{c}_{k+1} - \vec{c}_{j} \in {\mathbb L}$ implies
\beq
\vec{e}_k+\vec{c}_j +{\mathbb L}=\vec{e}_{k+1}+{\mathbb L} \;.
\eeq
Therefore we get
\beq
\theta_k( \vec{\phi} +  \tau \vec{c}_j)=e^{- \pi i \tau \vec{c}_j^2 -2\pi i \vec{c}_j \cdot \vec{\phi}} \theta_{k+1}(\vec{\phi} ) \;.
\eeq
Note that the overall factor $e^{- \pi i \tau \vec{c}_j^2 -2\pi i \vec{c}_j \cdot \vec{\phi}} $ is independent of $k$,
and hence is irrelevant when $\theta_k$ are mapped to homogeneous coordinates of $\CP^{N-1}$.

In summary, in terms of the homogeneous coordinates, the symmetries $ \bZ^{(A)}_N \times \bZ^{(B)}_N $ act as
\beq
&1 \in \bZ^{(A)}_N : [ \cdots, Z_k , \cdots] \mapsto [ \cdots, Z_{k+1} , \cdots]\;, \label{eq:centerA} \\
&1 \in \bZ^{(B)}_N : [ \cdots, Z_k , \cdots] \mapsto [ \cdots, e^{\frac{2\pi i k }{N}}Z_k , \cdots]\;, \label{eq:centerB}
\eeq
or more generally
\beq
(p,q) \in \bZ^{(A)}_N \times \bZ^{(B)}_N : [ \cdots, Z_k , \cdots] \mapsto [ \cdots, e^{\frac{2\pi i qk }{N}}Z_{k+p} , \cdots]\;.
\eeq
As expected, $\bZ^{(A)}_N$ and $\bZ^{(B)}_N$ commute with each other.

\section{\texorpdfstring{The $\CP^{N-1}$ instantons as Yang-Mills instantons}{The CP(N-1) instantons as Yang-Mills instantons}}
In this section we explain that the instantons of $\CP^{N-1}$-model can be identified with the Yang-Mills instantons.
First, we argue that their $\theta$ angles are identified, $\theta_{\CP^{N-1}}=\theta_{\rm YM}$.
Second, we argue the relation between one instanton solution of $\CP^{N-1}$-model and 
that of the Yang-Mills theory. Readers willing to accept the fact that [Yang-Mills instanton] = [$\CP^{N-1}$ instanton] can skip this section.
The relation between 4d instantons and 2d instantons are discussed in more mathematically precise way by Atiyah~\cite{Atiyah:1984tk} motivated by the analogy between
the Yang-Mills and the $\CP^{N-1}$ model.

\subsection{The topological theta term}
We explain the equivalence of the topological terms between the $\CP^{N-1}$ and Yang-Mills.
Actually, we discuss more details of the actions. 
We follow the explanation in section~4 of \cite{Kapustin:2006pk} adapted to our case.

\subsubsection{4d Yang-Mills as 2d gauge theory with infinite gauge group}
Let $E$ be a Riemann surface which for our application is $\bT^2$.
We consider a 4d spacetime $\bR^2 \times E$ with coordinates $(x^i, y^a)$, where $x^i~(i=0,1)$ are the coordinates of $\bR^2$ and 
$y^a~(a=2,3)$ are the coordinates of $E$. 
They are related to the complex coordinate $z$ on $E$ as $z=y^2+ i y^3$.
The metric on $E$ is denoted as $ds^2=g_{ab}dy^a dy^b =2 g_{z\bar{z}}|dz|^2$.
The following discussion can be easily generalized even if we replace $\bR^2 \times E$ by more general $X_2 \times \Sigma$ 
for 2d space-time $X_2$ and a Riemann surface $\Sigma$.

The 4d Yang-Mills theory with a compact, simple and simply connected gauge group $G$ on $\bR^2 \times E$ can be regarded as a 2d gauge theory on $\bR^2$
with gauge group ${\mathcal G}$, where ${\mathcal G}$ is the group of all gauge transformations on $E$.
Here we are regarding the space $E$ as an internal manifold, and considering a gauge theory on the 2d space-time $\bR^2$.
Notice that this ${\cal G}$ is an infinite dimensional group,
\beq
{\cal G}=\{ g: E \to G \}\;,
\eeq
with the obvious group structure.
In the following, all the Lie algebra generators are taken to be anti-hermitian which is different from the standard physics convention.

The total gauge field $A$ is split as $A=a+b$, where $a=\sum_{i=0,1}a_{i} dx^i$ is the gauge field of ${\cal G}$ and 
$b=\sum_{a=2,3} b_a dy^{a}$ is now regarded as ``matter fields" from the 2-dimensional point of view.
Regarding $b$ as matter fields may be intuitively understood if we perform Kaluza-Klein reduction,
since in the 2d space-time $\bR^2$ they are just scalar fields. However, we can proceed in an abstract way without any Kaluza-Klein reduction.

A gauge transformation $g(x,y)$ of $G$ on $\bR^2 \times E$ can be regarded as a gauge transformation of ${\cal G}$ on $\bR^2$, and it acts as
\beq
a \to g^{-1}a g+g^{-1} d_x g\;, \qquad b \to g^{-1}b g+g^{-1} d_y g\;,
\eeq
where $d_x$ and $d_y$ are exterior derivatives on $\bR^2$ and $E$, respectively.

The matter field $b$ takes values in the space of connections on $E$ which we denote as ${\cal A}$.
The space ${\cal A}$ is an infinite dimensional \Kahler manifold, once we fix a complex structure on $E$. The 
holomorphic coordinates on ${\cal A}$ are given by $b_{\bar z}$, i.e., the anti-holomorphic component of the connection 
$b=b_{\bar z} d \bar{z} + b_z dz$ on $E$.
The \Kahler form $\omega$ is given by\footnote{We change the normalization of the \Kahler form
$\omega$ from that of \cite{Kapustin:2006pk} by a factor of $2\pi$.}
\beq
2\pi \omega &= - \frac{1}{4\pi} \int_E \tr (\delta b\wedge \delta b) \nonumber \\
&= \frac{1}{2\pi} \int_E ( dz \wedge d\bar{z}) \tr (\delta b_{\bar z} \wedge \delta b_z )\;,  \label{eq:kahler}
\eeq
where $\delta$ is the exterior derivative on ${\cal A}$, or more intuitively, $\delta b$ means infinitesimal fluctuations of $b$.
The trace is normalized in such a way that the usual instanton number is given by integration of $\frac{1}{8\pi^2} \tr (F \wedge F)$.
(For the case of $G=\SU(N)$, the trace is in the defining $N$-dimensional representation.)
Even if we forget about the complex structure of $E$, the $\omega$ still defines a natural symplectic structure on ${\cal A}$.
Under the gauge transformation, the $\delta b$ transforms as $\delta b \to g^{-1} (\delta b) g$, and hence the gauge transformation preserves
the \Kahler form.

Because $\CG$ acts on $\CA$ preserving the \Kahler structure, we can define the moment map of $\CG$ as follows.
An element of the Lie algebra of $\CG$ is
an infinitesimal gauge transformation $\epsilon$ on $E$.
It is described by a vector field on $\CA$ given by
\beq
V(\epsilon) = \int_E ([b, \epsilon] + d_y \epsilon) \frac{\partial }{\partial b} \;,
\eeq
where $[b, \epsilon] + d_y \epsilon$ is the infinitesimal gauge transformation of $b$.
Let $\iota_V$ be the contraction operator of a vector field $V$ acting on differential forms. We get
\beq
\iota_{V(\epsilon)} (2\pi \omega) =  - \frac{1}{2\pi} \int_E \tr \left(    ([b, \epsilon] + d_y \epsilon)   \wedge   \delta b \right) 
=  \frac{1}{2\pi} \int_E \tr \left(    \epsilon   D_y \delta b \right)  \;,
\eeq
where $D_y=d_y +[b, \cdot]$ is the covariant exterior derivative.
Then we can write
\beq
\iota_{V(\epsilon)} (2\pi \omega) &= \delta \mu(\epsilon) \;,
\eeq
where $\mu(\epsilon) $ is the moment map associated to $\epsilon$, and is given by
\beq
\mu(\epsilon) &=  \frac{1}{2\pi} \int_E \tr \left(    \epsilon   f^b \right)  \;,\\
f^b &= d_y b+b \wedge b\;.
\eeq
We define $\mu$ by
\beq
\mu=\frac{\sqrt{g^{-1} } }{2\pi}f^b_{23}=-\frac{i g^{\bar{z}z}}{2\pi}f^b_{z\bar{z}}\;.
\eeq
This $\mu$ takes values in the Lie algebra of $\CG$.

Now we can rewrite the 4d Yang-Mills action as a $\CG$ gauge theory on the 2d spacetime $\bR^2$. 
First, let us introduce a notation for an inner product on the internal space.
For elements of the Lie algebra of $\CG$, $r$ and $s$, we define
\beq
\vev{ r, s} 
= - \int_E \sqrt{g}\, d^2y \tr (rs) \;.
\eeq
The minus sign comes from the fact that we take the Lie algebra to be anti-hermitian, and hence this inner product is positive definite.
Now the kinetic term can be written as
\beq
-\int_{\bR^2 \times E} \frac{1}{2g^2} \tr (F_{\mu\nu}F_{\mu\nu}) = \int_{\bR^2} L_{\rm 2d,kin} \;.
\eeq 
The Lagrangian $L_{\rm 2d,kin}$ is given by
\beq
L_{\rm 2d,kin} =  \frac{1}{g^2} \left[ \frac{1}{2}\vev{f^a_{ij}, f^a_{ij}} + 2\vev{D_i b_{\bar z}, g^{\bar{z}z} D_i b_{ z}}+ (2\pi)^2 \vev{\mu, \mu} \right] \;,
\eeq
where
\beq
D_i b_{\bar z} = \partial_i b_{\bar z} - \partial_{\bar{z}} a_i +[a_i, b_{\bar z}]\;, \qquad f^a_{ij}=\partial_i a_j - \partial_j a_i+[a_i, a_j]\;.
\eeq
Therefore, the 4d Yang-Mills is now rewritten as a 2d gauge theory with the gauge kinetic term $\vev{f^a_{ij}, f^a_{ij}}$, the matter kinetic term 
$\vev{D_i b_{\bar z}, g^{\bar{z}z}D_i b_{ z}}$ and the potential energy $\vev{\mu, \mu}$.

The topological term of 4d Yang-Mills is given as
\beq
\frac{i \theta}{8\pi^2} \int_{\bR^2 \times E}  \tr(F \wedge F) = \int_{\bR^2} L_{\rm 2d,top} \;,
\eeq
where
\beq
 L_{\rm 2d,top}=-\frac{i \theta}{4\pi^2}  \left[ 2\pi \vev{f^a, \mu} -i \vev{D_x b_{\bar z}, g^{\bar{z}z} D_x b_{z}} \right]\;,
\eeq
where $f^a=\frac{1}{2}f^a_{ij} dx^i \wedge dx^j$ and $D_x b_{\bar z}=D_i b_{\bar z} dx^i$.

\subsubsection{\texorpdfstring{$\CP^{N-1}$ as \Kahler quotient $\CA /\!/ \CG$}{CP(N-1) as Kahler quotient A/G}}
Having derived the 2d Lagrangian $L_{\rm 2d}=  L_{\rm 2d,kin}+L_{\rm 2d,top}$,
let us look at the potential minima of this Lagrangian. The potential energy is proportional to $\vev{\mu, \mu}$,
and hence the potential minima at the classical level is given by $2\pi \mu=f^b_{23}=0$.
This is just the flat connections on $E$. If we divide the space of the potential minima by the gauge group $\CG$,
we get the $\CP^{N-1}$ for the case $G=\SU(N)$ as discussed before. On the other hand, taking $\mu=0$ and dividing by the group $\CG$
is precisely the \Kahler (or symplectic) quotient of the space $\CA$ by $\CG$ denoted as $\CA/\!/\CG$.
Therefore we get
\beq
\CP^{N-1} = \CA/\!/\CG \;.
\eeq
The \Kahler form \eqref{eq:kahler} descends to this space $ \CA/\!/\CG$. 

Now let us look at the topological term in this subspace $\mu=0$. We get
\beq
 L_{\rm 2d,top} &\xrightarrow{\mu=0} \frac{i \theta}{2\pi}  \left[  \frac{i}{2\pi} \vev{D_x b_{\bar z}, g^{\bar{z}z}D_x b_{z}} \right] \nonumber \\
 &= \frac{i \theta}{2\pi}  \left[  \frac{1}{2\pi} \int (dz \wedge d \bar{z}) \tr(D_x b_{\bar z} D_x b_{z}) \right] \;.
\eeq
However, this is precisely the pull-back of the \Kahler form given by \eqref{eq:kahler}.
Let $\Phi$ be the map $\Phi: \bR^2 \ni x \mapsto b_{\bar{z}}(x) \in \CA$. Then, by the pull-back we get $\Phi^* \delta b_{\bar{z}} = D_x b_{\bar{z}}$,
and hence
\beq
\left[ \frac{1}{2\pi} \int (dz \wedge d \bar{z}) \tr (D_x b_{\bar z} D_x b_{z})  \right] = 2\pi \Phi^* \omega \;.
\eeq
Therefore, we finally get
\beq
 L_{\rm 2d,top} \xrightarrow{\mu=0} i \theta \Phi^* \omega \;.
\eeq

The $\omega$ is a \Kahler form on $\CP^{N-1} = \CA/\!/\CG$. 
Furthermore, it must be integrally quantized on $\CA/\!/\CG$
and can take the value $1$
as is clear from the fact that $ \Phi^* \omega$ comes from the 4d topological term $\frac{1}{8\pi^2} \int \tr (F \wedge F)$.
Therefore $\omega$ is the generator of $H^2(\CP^{N-1}, \bZ)$, and the instanton number of $\CP^{N-1}$ corresponds to the
instanton number of 4d Yang-Mills.

\subsection{Moduli space of instantons}
Next let us compare the moduli spaces of instantons of Yang-Mills and those of $\CP^{N-1}$-model.
To simplify the problem, we compactify the 2d space-time $\bR^2$ to $S^2$.

For the Yang-Mills theory, the total 4d space-time is $S^2 \times E$.
Then, the Atiyah-Hitchin-Singer theorem~\cite{Atiyah:1978wi} states that the number of (virtual\footnote{Virtual dimensions coincides with the actual
dimensions for irreducible solutions in which the gauge group is completely broken by the solutions.}) 
dimensions of the instanton moduli space
for $\SU(N)$ with topological charge $Q$ is given by
\beq
4N Q - (N^2-1)\frac{\chi + \sigma}{2} \;,
\eeq
where $\chi$ and $\sigma$ are the Euler number and the signature of the 4-manifold.\footnote{Small fluctuations of gauge fields $\delta A$
are described as $\delta A^a_{\alpha \dot{\beta}}$, where $\alpha$ and $\dot{\beta}$ are spinor indices of Lorentz symmetry
$\SO(4) \cong \SU(2)_L \times \SU(2)_R$, and $a$ is the index of the adjoint representation of $\SU(N)$. 
By regarding it as an $\SU(2)_L$ spinor field $\delta A^a_{\alpha \dot{\beta}}=(\lambda_\alpha)^a_{\dot{\beta}}$ which takes values in the $\SU(N) \times \SU(2)_R$ bundle,
one can apply the usual Atiyah-Singer index theorem to get the Atiyah-Hitchin-Singer index theorem. }
For the 4-manifold $S^2 \times E$,
we have $\chi=\sigma=0$, and thus the dimensions is $4NQ$.

Let us next consider the instantons of the $\CP^{N-1}$-model on $S^2$. If we regard $S^2=\CP^1$,
then the instantons of $\CP^{N-1}$-model is given by a holomorphic map $\CP^1 \to \CP^{N-1}$,
and the instanton number is given by the winding number of this map.
The standard explanation is as follows. Consider a general sigma model whose target space $\CM$ is K\"{a}hler. 
Let $\xi^I$ be holomorphic coordinates of the sigma
model, and let $g_{I\bar{J}}$ be the \Kahler metric such that the \Kahler form $\omega=i g_{I\bar{J}}  d\xi^I \wedge d\bar{\xi}^{\bar{J}}$ is integrally 
quantized, i.e., the generator of $H^2(\CM, \bZ)$.
Also let $x^i$ be the coordinates of the spacetime $S^2=\CP^1$ with $w=x^0+i x^1$ its complex coordinate. Then we get
\beq
 & g_{I\bar{J}}(\partial_i \xi^I \pm \epsilon_{ij} \partial_j \xi^I)(\partial_i \bar{\xi}^{\bar{J}} \pm \epsilon_{ij} \partial_J \bar{\xi}^{\bar{J}} ) \geq 0 \nonumber \\
\Rightarrow ~& \int d^2x g_{I\bar{J}}(\partial_i \xi^I )(\partial_i \bar{\xi}^{\bar{J}}  ) \geq 
\left| \int d^2x g_{I\bar{J}}  \epsilon_{ij} (\partial_i \xi^I )(\partial_j \bar{\xi}^{\bar{J}}  ) \right| =\left| \int \omega \right|=|Q| \;,
\eeq
where $Q$ is the instanton number defined in terms of the \Kahler form $\omega$.
The equality is saturated if and only if $\partial_i \xi^I \pm \epsilon_{ij} \partial_j \xi^I=0$, which implies
$\partial_{\bar{w}} \xi^I =0$ (instanton) or $\partial_{{w}} \xi^I =0$ (anti-instanton) in terms of $w=x^0+i x^1$.
The equation $\partial_{\bar{w}} \xi^I =0$ means that the map from the space-time $\CP^1$ to the target space $\CM$ is holomorphic.

Here we consider the case where the instanton number is integer. The case of fractional instantons is discussed in section~\ref{sec:twisted}.

Let $[ Z_k]$ ($k=1,\cdots,N$) be the homogeneous coordinates of $\CP^{N-1}$,
and let $w \in \bC \cup \{ \infty \}=\CP^1$ be the usual coordinate of $\CP^1$. Then, holomorphic maps are given by
\beq
Z_k = \sum_{\ell=0}^Q a_{k,\ell} w^\ell \;,
\eeq
where $a_{k,\ell}$ are constants. The fact that the degree of the polynomial $Q$ corresponds to the instanton number
is easy to see by considering simple cases. Let us consider the case $N=2$ and $[Z_1,Z_2]=[1,w]$.
This map is the identity map from the spacetime $\CP^1$ to the target space $\CP^{N-1=1}$. Therefore it gives the instanton number $Q=1$.
In the case $[Z_1,Z_2]=[1,w^Q]$, the spacetime $\CP^1$ wraps $Q$ times around the target $\CP^{N-1}$, so the instanton number is $Q$.
The general case is seen by using the invariance of the topological charge under continuous deformation, 
and also by using the embedding $\CP^1 \to \CP^{N-1}$ as $[Z_1,Z_2] \to [Z_1,Z_2,0,\cdots,0]$.

There are $N(Q+1)$ complex parameters $a_{k,\ell}$. However, the overall scale is irrelevant 
because $Z_k$ are homogeneous coordinates. So there are $N(Q+1)-1$ complex parameters, or real $2N(Q+1)-2$ parameters.

When $Q=1$, the $\CP^{N-1}$ instantons have real $4N-2$ parameters. Thus there are two missing parameters compared with 
the dimension of the Yang-Mills instantons $4N$ discussed above. 
These missing moduli may be understood as follows. On generic points of the moduli space $\CM_{\rm flat}=\CP^{N-1}$,
the $\SU(N)$ gauge group is broken down to $\U(1)^{N-1}$. However, along some locus on $\CM_{\rm flat}$, some non-abelian gauge symmetry
is recovered.
The $\CP^{N-1}$-model is an approximate description of the Yang-Mills theory away from the locus where non-abelian symmetries are enhanced.
When $\CP^{N-1}$ instantons hit such a locus, the approximation in terms of the flat connection \eqref{eq:flat} is not valid and 
the gauge field develops nontrivial profile along $E$~\cite{Friedman:1997yq}. 
Then the full Yang-Mills solution may not be invariant under the translations along the torus $E$.
The moduli about translations along $E$ are not visible in the low energy $\CP^{N-1}$ description. Since $E$ has 2 directions,
there are 2 translation moduli. This provides the missing 2 moduli parameters.

When $Q>1$, the number of the parameters of instantons of $\CP^{N-1}$-model are much smaller than those of the Yang-Mills theory. The missing moduli parameters can still be described in algebraic geometry by using spectral cover construction~\cite{Friedman:1997yq}
because $\CP^1 \times E$ can be regarded as a (trivial) elliptic manifold.
We do not discuss it in this paper, however.

\section{Fractional instantons, vacuum structure, and center symmetry}\label{sec:twisted}

Having studied the relation between instantons in 4d and 2d theories, we now study fractional instantons.
In particular we show that such instantons restore the center symmetry of the system.

Let $(t,x)$ be the coordinates of $ \bR \times S^1_C$ with $x \sim x+1$.
The metric may then be taken to be
\beq
ds^2= L_C^2(dt^2+dx^2)+L_E^2|dz|^2,
\eeq
where $L_C$ and $L_E$ are some length parameters. The multiplication of the term $dt^2$ by the same $L_C^2$ as the one
multiplying $dx^2$ is introduced just for later convenience. 

For the twisted compactification, we choose to use the symmetry $\bZ^{(B)}_N$ discussed in Sec.~\ref{sec:oneform}.
Suppose we first compactify the theory on $E$ and reduce it to $\CP^{N-1}$. 
Then we learn from  \eqref{eq:centerB} that the twist by $\bZ^{(B)}_N$ on the homogeneous coordinates is 
given by
\beq
[\cdots, Z_k(t, x+1),\cdots] = [\cdots, e^{\frac{2\pi i k }{N}} Z_k(t,x),\cdots]  \;.\label{eq:twisted}
\eeq
This is precisely the boundary condition which played an important role in the analysis of the $\CP^{N-1}$ 
model~\cite{Dunne:2012ae,Dunne:2012zk,Sulejmanpasic:2016llc}.
For this reason let us start our analysis in the $\mathbb{CP}^{N-1}$-model, and then subquently discuss 
the four-dimensional theory.

\subsection{\texorpdfstring{Fractional instantons and the vacuum structure in $\mathbb{CP}^{N-1}$-model}{Fractional instantons and the vacuum structure in CP(N-1)-model }}
Here we discuss the vacuum structure of the $\CP^{N-1}$-model. 
First of all, let us discuss the candidates for the vacuum at the classical level in the presence of the twist \eqref{eq:twisted}.
A minimal energy configuration should be constant on 2d space-time $(t,x)$, and also invariant under the above twist.
This requires $ e^{2\pi i k /N} Z_k = c Z_k~(k=1,\cdots,N)$, where $c$ is a constant independent of $k$ which takes care of the fact that $Z_k$ are
homogeneous coordinates and hence an overall constant is irrelevant.
There are $N$ solutions of these equations, which are given by
\beq
P_k=[0,\cdots, 0, \overset{k}{1},0,\cdots,0] \in \CP^{N-1}. \label{eq:candidatevac}
\eeq
These points are the fixed points of the twisting \eqref{eq:twisted}. These are the classical vacua.

Thanks to the twisting, the fields have energy gap. For example, near the point $P_N$, the fields are described
by inhomogeneous coordinates $U_k :=Z_k/Z_N$ with the boundary condition $U_k(t,x+1)=e^{\frac{2\pi i k}{N}}U_k(t,x)$.
This boundary condition forces $U_k$ to have a nonzero momentum in the direction $S^1_C$.
Therefore, the vacua have energy gap already at the classical level, so there are no IR divergences.
This makes straightforward perturbation theory possible around these vacua.

There are transitions among these classical vacua $P_k$ by fractional instantons and anti-instantons, as we now describe. 
First, let us identify $\bR \times S^1_C$ as $S^2=\CP^1$ by regarding the infinite future $t=\infty$ and infinite past $t=-\infty$
as the north pole and south pole of $S^2=\CP^1$, respectively. More explicitly, we take the coordinate
\beq
w = e^{2\pi (t+ix)} \in \bC+\{0\}=\CP^1.
\eeq
The point $w=0$ corresponds to the infinite past $t=-\infty$ and the point $w=\infty$ corresponds to the infinite future $t=+\infty$.
Then, as stated in the previous section, instantons are given by holomorphic maps from $\CP^1$ to $\CP^{N-1}$.
However, we have to take into account the boundary condition \eqref{eq:twisted} appropriately when we consider instanton solutions.

\begin{figure}[htbp]
\centering\includegraphics[scale=0.45]{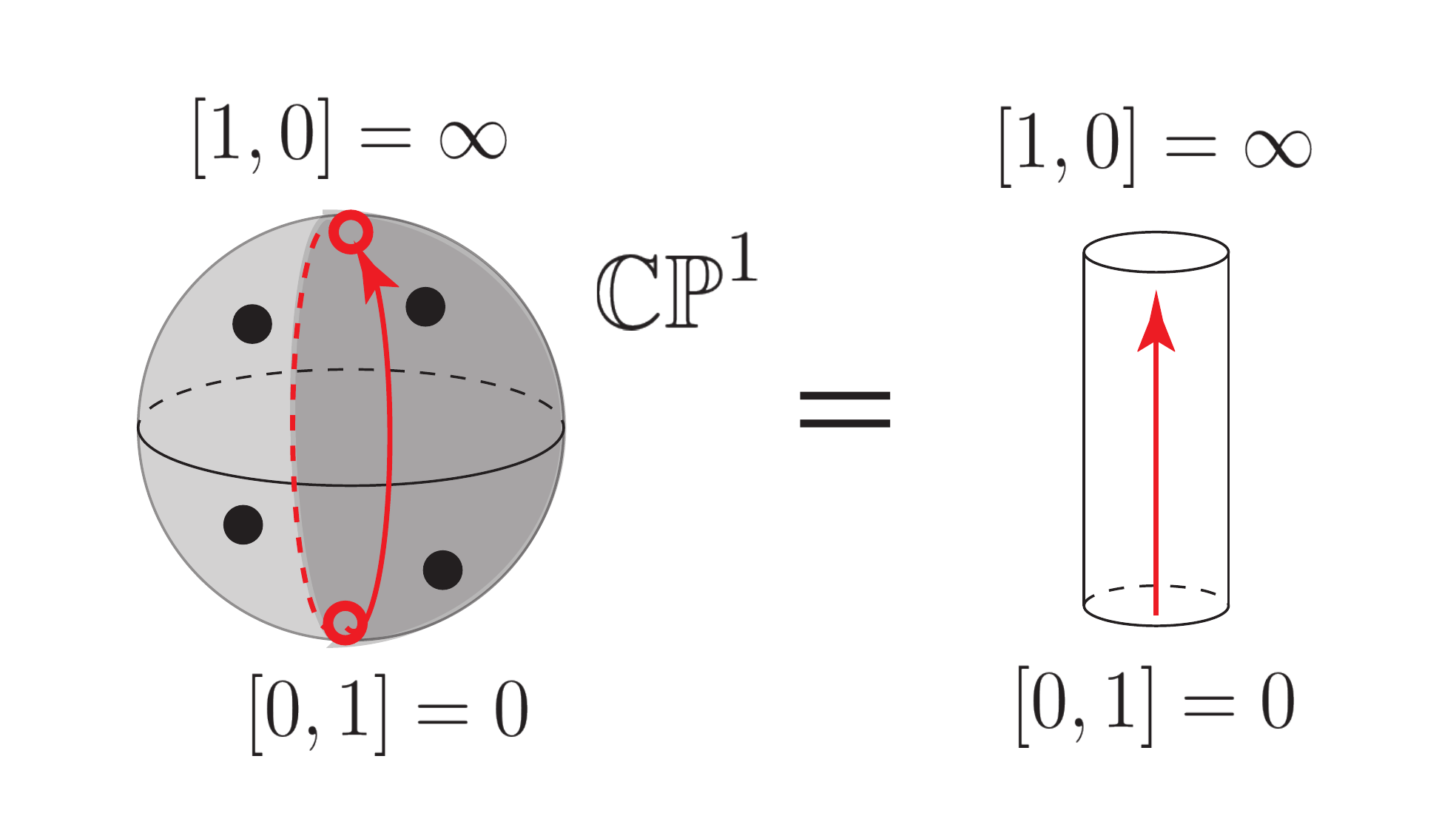}
\caption{A fractional instanton for $\mathbb{CP}^1$-model interpolates between  
two different vacua, $0$ and $\infty$ (the black dots represents singular points, as discussed in Figure \ref{fig:T4Z2}). In cylindrical coordinate (so that the geometry is $\mathbb{R}\times S^1$) with fixed $x$, this 
moves along the $\mathbb{R}$-direction from past to future infinities, as we vary the parameter $t$. The full trajectory, with both $t$ and $x$ varied, is a hemisphere, and when the two such hemispheres are combined we obtain a full sphere.
This is a manifestation of the fact that two fractional instantons combine into a single instanton.}
\label{fig:CP1fractional}
\end{figure}

\begin{figure}[htbp]
\centering\includegraphics[scale=0.45]{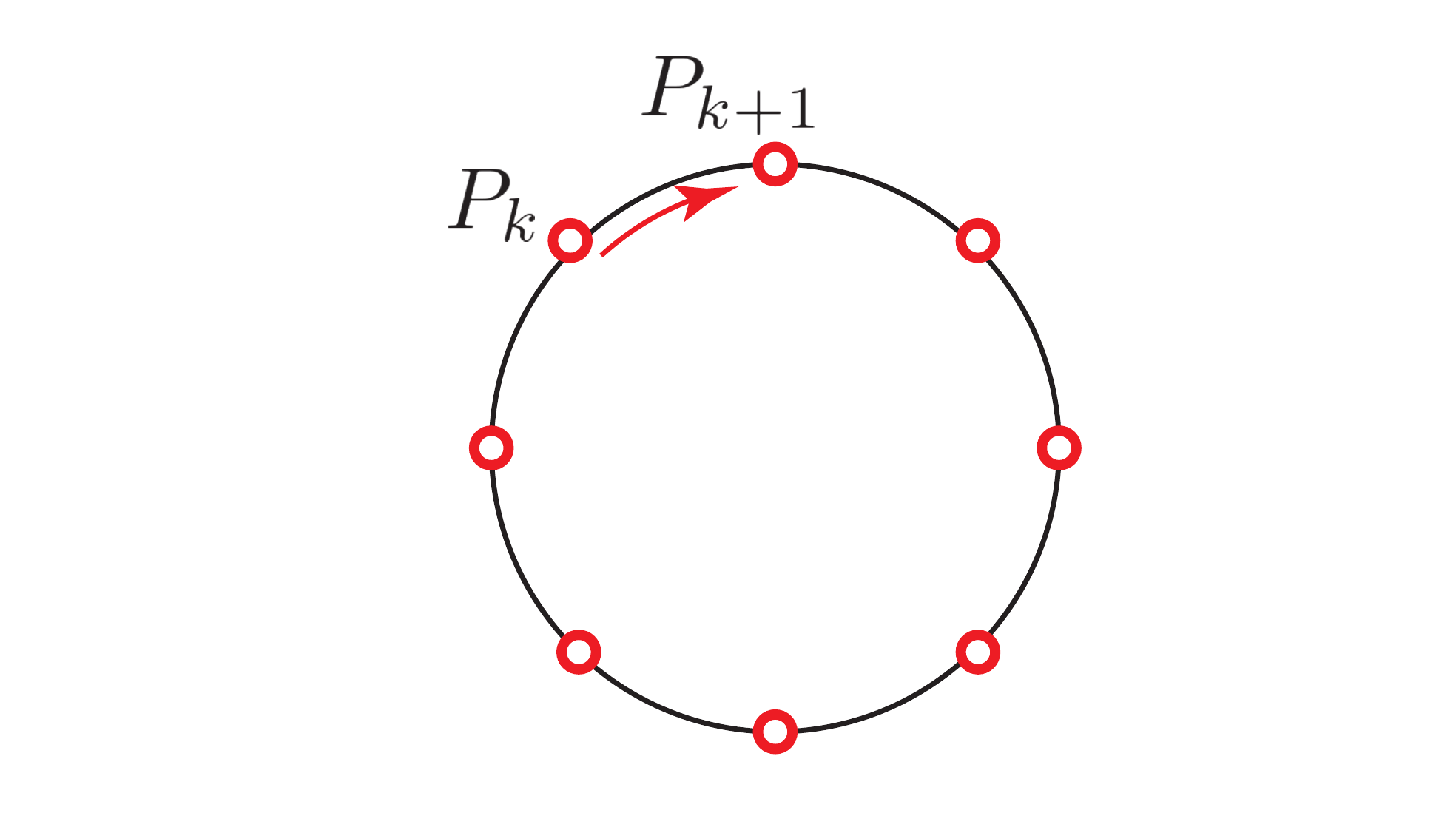}
\caption{The geometry is more complicated for $N>2$ case than the $N=2$ case in Figure \ref{fig:CP1fractional}.
However, the idea is the same: the classical moduli space $\mathbb{CP}^{N-1}$ is lifted by twisted boundary condition along $S^1_C$ into $N$
different points  $P_1, \ldots P_N$, however tunneling between different these points, as given by fractional instantons, 
lifts the $N$-fold degeneracy, restoring the center symmetry dynamically.
}
\label{fig:CPNfractional}
\end{figure}

In the present case of twisted compactification, there exist fractional instantons (see \cite{Eto:2004rz,Eto:2006mz,Bruckmann:2007zh,Brendel:2009mp,Harland:2009mf} for early works). 
There are $1/N$ instantons connecting $P_k$ and $P_{k+1}$ which we denote as $I_{\khalf}$.  
They are shown in Figure~\ref{fig:CP1fractional} and \ref{fig:CPNfractional} and given by
\beq
I_{\khalf}= & (P_k \to P_{k+1}):~~~[\cdots,Z_k(w),Z_{k+1}(w),\cdots]=[0,\cdots, 0,a,w^{\frac{1}{N}},0,\cdots,0]  \;,
\eeq
where $a$ is a constant complex moduli parameter of the $1/N$ instantons. 
Notice that the twisted boundary condition \eqref{eq:twisted} is satisfied by these fractional instanton solutions. 
The loop $x \to x+1$ corresponds to $w \to e^{2\pi i}w$,
under which $w^{1/N} \to e^{2\pi i /N} w^{1/N}$. By using the fact that homogeneous coordinates $Z_k$ are defined only up to overall multiplication by
constants, one can check that \eqref{eq:twisted} is satisfied. 
This really has $1/N$ topological charge, because by going to the $N$-covering space of $\CP^1$ as $w'=w^{1/N}$,
the above solutions are $[\cdots,0,a,w',0,\cdots]$ which is the standard one instanton of $\CP^{N-1}$.
There are also $-1/N$ anti-instantons $\bar{I}_{\khalf}$ which are obtained by replacing $w \to 1/\bar{w}$.

Note that the Atiyah-Hitchin-Singer theorem~\cite{Atiyah:1978wi} 
predicts that there are $4NQ=4$ moduli fields for $Q=1/N$ instantons. Two of the four real parameters are provided by $a$.
The other two may be provided by the moduli associated to translations along $E$ as discussed earlier in the case of $Q=1$.
Actually, the $\frac{1}{2\pi i}\log a$ also gives translation moduli along the direction $\bR \times S^1_C$.
Therefore, all the four moduli parameters of the Yang-Mills $1/N$ instantons are the translations on $\bR \times \bT^3$.\footnote{
For some gauge groups $G$ other than $\SU(N)$, there exist fractional instantons on $\bR \times \bT^3$ even without twisting.
In those cases, the fact that the four moduli parameters are given by translations on $\bR \times \bT^3$ can be seen by
using string duality. See section 3.1 of \cite{Ohmori:2015pua} for these discussions.}

Now the true vacuum structure of the theory at the quantum level is described as follows. 
First of all, note that in the present case, the $1/N$ instantons do not have size modulus because all
the four moduli are associated to translations on space-time $\bR \times  \bT^3$. 
The absence of the size modulus is related to the fact that there are no zero modes in the twisted compactification used in this paper.
Then, at least if the length scale of the compactification is small,
the dilute gas approximation of fractional instanton/anti-instanton is valid. 
This is the crucial difference from the instanton computation on $\bR^4$ discussed in the Introduction.

Let $H$ be the Hamiltonian of the system, and $\ket{P_k}$ be the classical vacuum staying at the point $P_k$.
Then the usual dilute gas computation of instantons
would give the transition amplitudes
\beq
\bra{P_{k+n}} H \ket{P_k} = C_n(g,e^{i\theta},L_A,L_B,L_C, \mu) \, e^{\frac{i n\theta }{N}}\;, \label{eq:transition}
\eeq
where $C_n(g,e^{i\theta},L_A,L_B,L_C, \mu)$ is a function of the coupling $g$, the exponential of the theta angle $e^{i\theta}$,
the lengths $L_{A,B,C}$ along $S^1_{A,B,C}$ and the renormalization scale $\mu$. 
The $k$ and $n$ should be considered mod~$N$ by regarding $C_{n+N} :=C_n e^{-i\theta}$.
This $C_n(g,e^{i\theta},L_A,L_B,L_C, \mu)$ has perturbative as well as non-perturbative terms, and is expected to have the form of a resurgent function.
The dependence of $C_n(g,e^{i\theta},L_A,L_B,L_C, \mu)$ on the theta angle $\theta$ through $e^{i\theta}$ is because
one-instanton $\prod_{k=1}^N I_{\khalf}$ is a transition from one vacuum to itself. 
In the dilute gas approximation, we roughly get
\beq
C_{n = \pm 1}(g,e^{i\theta},L_A,L_B,L_C, \mu) \sim -M\exp\left( -\frac{8\pi^2}{ Ng^2}  \right) \;,
\eeq
where $M>0$ is given by a combination of the parameters $(g,e^{i\theta},L_A,L_B,L_C, \mu)$ with mass dimension one,
while we neglect other $C_{n}(g,e^{i\theta},L_A,L_B,L_C, \mu)$ for $|n| \geq 2$.
The minus sign was put to make the later results consistent with the general argument of Vafa and Witten~\cite{Vafa:1984xg},
as was already mentioned around \eqref{eq:E_weak}.

By diagonalizing the Hamiltonian, we get the energy eigenstates and eigenvalues as
\beq
  H\ket{\el } =E_\el \ket{\el }\;, \qquad
   \ket{\el } = \frac{1}{\sqrt{N}}\sum_{k=1}^N e^{\frac{2\pi i k \el}{N}}\ket{P_k}\;,\label{eq:truevacua}
\eeq
where
\beq
E_\el &= \sum_{n=0}^{N-1}C_n(g,e^{i\theta},L_A,L_B,L_C, \mu) \, e^{\frac{i n(\theta - 2\pi \el)}{N}} \\
& \sim C_0-2M \exp \left( -\frac{8\pi^2}{ Ng^2}  \right) \cos \left(  \frac{\theta-2\pi \el}{N}  \right)\;.
\eeq
There are $N$ states, and the true vacuum is the one which minimizes this energy 
while the others are metastable.


Let us notice several important properties of the above vacuum structure. 
First, the energies $E_\el$ depend on $e^{i \theta/N}$, while the $\theta$ is supposed to be a periodic variable $\theta \sim \theta + 2\pi$.
They are consistent due to the existence of $N$ (meta)stable vacua. For each $\el \in \bZ_N$, we have one vacuum.
When the $\theta$ is adiabatically changed to $\theta+2\pi$, the vacuum $\ket{\el}$ goes to the vacuum $\ket{\el -1}$,
mapping one vacuum to another. This turns a metastable vacuum to a true vacuum, or a true vacuum into a metastable vacuum.
This is a manifestation of the Witten effect \cite{Witten:1979ey}, because our twisted boundary condition by the center symmetry
is equivalent to including a unit 't~Hooft magnetic flux ${\mathsf m}=1$ on $S^1_B \times S^1_C$~\cite{tHooft:1979rtg}, and the 
$\el \in \bZ_N$ can be identified with the 't~Hooft electric flux, as will be seen in later discussions.
In this way, the dependence on $\theta$ as $e^{i \theta/N}$ is consistent with the periodicity.
See \cite{Gaiotto:2017yup} for recent related discussion of this type of Witten effect.

Next, in the large $N$ limit and with finite $\el$, the energy is proportional to
\beq
E_\el \propto  (\theta - 2\pi \el)^2 \qquad (N \to \infty)\;.
\eeq
These properties are exactly as expected from the large 
$N$ analysis of $\CP^{N-1}$-model and Yang-Mills theory, as mentioned already in section \ref{subsec:IR}.

The results above are obtained in a small volume region where weak coupling analysis is reliable. However,
the above vacuum structure such as $E_\el \propto \cos [(\theta -2\pi \el)/N]$ was also obtained
in a different regime where the volume of the space-time is infinite while there is a light adjoint Majorana fermion~\cite{Yonekura:2014oja}.
The dependence of $E_\el$ on $\theta$ can have important implications for cosmology of (non-QCD) axions such as string axions, and hence it would be very interesting to investigate it further.

\subsection{4d situation}
The discussion of the previous subsection was for the $\CP^{N-1}$ model.
Let us see what is happening in the 4d Yang-Mills theory. 
We introduce coordinates $(x_A,x_B,x_C)$ related to the previously introduced coordinates as $z=x_A+\tau x_B $ and $x=x_C$ (where $\tau=iL_B/L_A$),
and call the corresponding directions as A, B and C cycles, respectively.

We would like to identify the points $P_k$ given in \eqref{eq:candidatevac}
in terms of the 4d Wilson lines $U_A$, $U_B$, and $U_C$, where these Wilson lines are computed along the 
A, B and C cycles. In the following, it may be convenient to consider the covering space $\bR^3$ of $S^1_A \times S^1_B \times S^1_C$
and regard the gauge fields as living on this $\bR^3$ with certain periodic properties.
The Wilson lines $U_{A,B,C}$ are defined by integration over $0 \leq x_{A,B,C} \leq 1$ as
\beq
U_{A}(x_B,x_C) = \exp \left( - \int_{0 \leq x_A \leq 1, \textrm{with fixed } x_B,x_C}  A \right) \;,
\eeq
and similarly for $U_B$ and $U_C$.

Implementation of the twist is a little more complicated in the Yang-Mills \cite{tHooft:1979rtg}. 
(See \cite{Witten:2000nv} for a detailed explanation of mathematical background behind the following discussion.)
We introduce a gauge transformation $h(x_B)$ which has the property that 
\beq
h(x_B+1)=e^{\frac{2\pi i}{N}}h(x_B) \;, \qquad h(x_B=0)=1 \;.
\eeq 
Then, the gauge field is assumed to have the periodic boundary condition
\begin{align}
\begin{split}
A(x_B,x_C+1) &=h^{-1}(x_B) A(x_B,x_C) h(x_B) + h^{-1}(x_B) d h(x_B) \;,\\
A(x_B+1,x_C) &= A(x_B,x_C) \;, \label{eq:thoofttwist}
\end{split}
\end{align}
where we suppressed $x_A$ because it is irrelevant for the discussion of the nontrivial part of the periodic boundary condition.
This periodicity property means in particular that
\begin{align}
U_B(x_C+1) &= h(x_B=1)^{-1} U_B(x_C)h(x_B=0) \nonumber \\
&=e^{-\frac{2\pi i}{N}}U_B(x_C)\;.
\end{align}
This corresponds to the twisted boundary condition $Z_k(x_C+1)=e^{2\pi i k/N}Z_k(x_C)$ of the $\CP^{N-1}$
as can be seen from \eqref{eq:YM1form} and \eqref{eq:centerB}.

Another point which requires a care is the following. Under gauge transformation $g$ on the space-time,
the $U_C$ transforms as $g^{-1}(x_B,x_C=1) U_C g(x_B,x_C=0)$. However, the $g$ also satisfies the boundary condition
$g(x_B,x_C+1)=h^{-1}(x_B) g(x_B,x_C)h(x_B)$ and hence the gauge tranformation is
$U_C \to h^{-1}(x_B) g^{-1}(x_B,x_C=0) h(x_B) U_C g(x_B,x_C=0)$.
This requires us to modify the definition of $U_C$ as
\beq
U'_C(x_B) = h(x_B)U_C(x_B)\;.
\eeq
Then the gauge invariant Wilson loops can be defined as $\tr U'_C$.
This new $U'_C$ has a boundary condition as
\beq
U'_C(x_B+1)=e^{\frac{2\pi i}{N}} U'_C(x_B)\;.
\eeq
The reason for this is as follows. As mentioned in section~\ref{sec:oneform} and reviewed in appendix~\ref{app:one_form}, 
the symmetry operators of the 1-form symmetry
$C(Y)$ is defined on some codimension-2 surface $Y$. 
One way to implement the twisted boundary condition (which might look different from
the implementation by \eqref{eq:thoofttwist} but is actually equivalent) is by inserting such an operator.
In the present case, the $Y$ extends in the direction
$\bR \times S_A  $ while it is localized on a point $\text{pt} \in S_B \times S_C$, so $Y=\text{pt} \times \bR \times S_A $. 
Schematically we write this twist as $C(\bR \times S_A )$.
Then the directions $S_B$ and $S_C$ are on the same footing in this implementation of the 1-form twist.
In particular, when we move $U'_C$ from $x_B$ to $x_B+1$, this Wilson line is crossed across the operator $C( \bR \times S_A)$.
This crossing produces the phase factor $e^{2\pi i/N} $ in the boundary condition of $U'_C$.
Similarly $U_B$ gets the phase $U_B(x_C+1) =e^{-\frac{2\pi i}{N}}U_B(x_C)$ by crossing $C( \bR \times S_A)$.
However, instead of $C( \bR \times S_A)$, we just use the boundary conditions \eqref{eq:thoofttwist} in this subsection.\footnote{Mathematically speaking,
any realization of the twist is described by a nontrivial topology of the $\SU(N)/\bZ_N$ bundle over $S^1_B \times S^1_C$. 
The \eqref{eq:thoofttwist} is one way to realize such a nontrivial bundle. Another way to realize it is to consider a small patch near 
$\text{pt} \in S^1_B \times S^1_C$, and take a nontrivial transition function around this point by using an element of $\pi_1(\SU(N)/\bZ_N)$. 
These two are equivalent.}

Now let us restrict our attention to flat connections
$F=dA+A^2=0$. In this case we have
\beq
U_C(x_B+1)  U_B(x_C) U^{-1}_C(x_B)  U^{-1}_B(x_C+1) \xrightarrow{\text{flat connection}} 1 \;.
\eeq
By using $U_C(x_B+1)=U_C(x_B)=h^{-1}(x_B)U'_C(x_B)$ and setting $U_{B}:=U_{B}(0)$ and $U'_C:=U'_C(0)$, we get
\beq
U'_C U_B =e^{-\frac{2\pi i}{N}}U_B  U'_C\;. \label{eq:noncomuute}
\eeq
More generally, if $U'_C U_B =e^{-\frac{2\pi i {\mathsf m}}{N}}U_B  U'_C$, then there is a 't~Hooft magnetic flux ${\mathsf m} \in \bZ_N$
on $S^1_B \times S^1_C$ \cite{tHooft:1979rtg,Witten:2000nv}.

Up to conjugation (i.e., gauge transformation), there is a unique solution to the equation \eqref{eq:noncomuute}.
This is because $U'_C$ can be regarded as the ``raising operator" of the eigenvalues of $U_B$. 
The solution up to conjugation is given by
\beq
(U_B)_{ij}= \delta_{i, j} \exp \left( 2 \pi i \frac{j-\frac{N+1}{2}}{N} \right),~~~~~(U'_C)_{i j}=\delta_{i,j+1}\;.
\eeq
On the other hand, $U_A=U_A(0)$ satisfies
\beq
U_AU_B=U_BU_A\;, \qquad U_AU'_C=U'_CU_A\;,
\eeq
because the boundary condition in the direction $x_A$ is trivial (i.e., there is no twist in the $S_A$ direction).
Then, there are $N$ solutions for $U_A$ given by
\beq
U_A = e^{2\pi i \frac{K}{N}} \cdot 1\;,
\eeq
where $K \in \bZ_N$. 

Notice that the gauge symmetry is completely broken by the Wilson lines $U_B$ and $U'_C$, because the subset of $\SU(N)$ whose elements commute
with both $U_B$ and $U'_C$ is discrete.
This makes all fields massive and there is no infrared divergence, and we are 
away from the singular points of the moduli space of flat connections on the two-torus $E$ because W-bosons are massive.

The above Wilson lines correspond to the coordinates of $\CP^{N-1}$ as follows.
In terms of $\vec{\phi}$ introduced in \eqref{eq:Wloop}, the above $(U_A,U_B)$ are given by
\beq
\vec{\phi} =  \frac{\vec{\rho}}{N} -\tau K \vec{c}_j\;,
\eeq
where $\vec{c}_j$ was introduced in \eqref{eq:centervector}, and $\vec{\rho}$ is 
 the Weyl vector (i.e., half the sum of all positive roots)  of $\SU(N)$ with respect to the root lattice 
${\mathbb L}$ \eqref{eq:root}:
\beq
\vec{\rho} = \left( \frac{N-1}{2}, \frac{N-3}{2}, \cdots, -\frac{N-1}{2}\right)\;.
\eeq

Now let us map the above $\vec{\phi}$ to $\CP^{N-1}$. First, by a computation as in Sec.~\ref{sec:oneform} we get
\beq
\theta_k\left( \frac{\vec{\rho}}{N} -  \tau K \vec{c}_j \right)=e^{- \pi i \tau K^2 \vec{c}_j^2 +2\pi i K \vec{c}_j \cdot \frac{\vec{\rho}}{N}} \theta_{k-K}\left(\frac{\vec{\rho}}{N} \right) \;.
\eeq
Next, let $W \in {\mathfrak S}_N$ be an element of the Weyl group such that it sends the $i$-th component to $i+1$-th component. 
Then, one can see that $W (\vec{\rho}) =\vec{\rho}-N\vec{c}_1$.
Noticing the fact that the lattices $\vec{e}_k+{\mathbb L}$ are invariant under the Weyl group, we get
\beq
& \theta_{k}\left(\frac{\vec{\rho}}{N} \right) = \theta_{k}\left(W\left(\frac{\vec{\rho}}{N}\right) \right) =  \theta_{k}\left(\frac{\vec{\rho}}{N} -\vec{c}_1\right)=e^{\frac{2\pi i k}{N}} \theta_{k}\left(\frac{\vec{\rho}}{N} \right)\;,
\eeq
and hence $ \theta_{k}(\vec{\rho}/N )=0$ for $k \neq 0 \mod N$. 
Therefore, we conclude that
\beq
\theta_k\left( \frac{\vec{\rho}}{N} -  \tau K \vec{c}_j\right)=0 ~~\text{for}~~k \neq K\;.
\eeq
By the identification $Z_k=\theta_k$, this is precisely the point $P_K$ introduced in \eqref{eq:candidatevac};
\beq
U_A = e^{\frac{2\pi i K}{N}}  \longleftrightarrow P_K=[\cdots, 0, \overset{K}{1},0,\cdots] \;.
\eeq

\subsection{Dynamical restoration of the center symmetry}\label{sec:center_restore}
Now we can show that the center symmetry (i.e.\ dimensional reduction of 1-form symmetry) is unbroken. 
Because we are compactifying the theory on $S^1_A \times S^1_B \times S^1_C$, we get three center symmetries
$\bZ^{(A)}_N \times \bZ^{(B)}_N \times \bZ^{(C)}_N$ from 
the dimensional reduction of 1-form symmetry, and they are
associated to the gauge invariant Wilson loop operators $\tr U_A$, $\tr U_B$ and $\tr U'_C$.
For operators $\tr U_B$ and $\tr U'_C$, we have
\beq
\tr U_B = \tr U'_C=0 \;.
\eeq
Therefore, the only nontrivial point is about the operator $\tr U_A$ which acts on $\ket{P_k}$ as
\beq
 \tr U_A \ket{P_k}= N e^{\frac{2\pi i k}{N}}  \ket{P_k} \;.
\eeq
Thus, at each classical vacuum on $P_k$, the center symmetry is broken by the VEV (vacuum expectation value) of the Wilson loop $\tr U_A$.
However, we found that the true quantum vacua are linear combinations of $\ket{P_k}$ given in \eqref{eq:truevacua}.
For them, we get
\beq
\bra{\el} \tr U_A \ket{\el} = \frac{1}{N} \sum_{k=1}^N N e^{\frac{2\pi i k}{N}} = 0 \;. \label{eq:wilAonP}
\eeq
Therefore, the center symmetry is dynamically restored by $1/N$ fractional instanton/anti-instanton effects!
The fact that the symmetry is restored itself is not so surprising because all the spatial directions are compact $\bT^3=S^1_A \times S^1_B \times S^1_C$
and a symmetry breaking does not usually happen in quantum mechanics. However, we stress that 
the ``strong dynamics" properties of confinement such as the center symmetry restoration and the dependence of the vacuum energy on $\theta$
are realized completely in weak coupling regime.
Because of this, we conjecture that the vacua discussed in the previous subsection are adiabatically continued 
when we go from small to large volume limit.

Although the center symmetry is unbroken, the vacua have charges under $\bZ^{(A)}_N$.
The vacuum states go to itself under the action of the symmetry 
and hence all operators having nonzero charge under it have vanishing VEVs. 
But there can be a possible phase factor on the action of the symmetry operator on the vacua.
To see this, let $C_A$ be the generator of the center symmetry $\bZ^{(A)}$ acting on the Hilbert space.\footnote{
In the description $C(Y)$ in appendix~\ref{app:one_form}, $C_A$ is defined by choosing $Y$ which is extending in the direction $S_B \times S_C$
while localized in the direction $\bR \times S_A$. So schematically we can write $C_A = C(S_B \times S_C)$. }
From \eqref{eq:centerA}, it acts as $C_A\ket{P_k} =  \ket{P_{k-1}}$.\footnote{In principle, 
we can have a nontrivial phase factors here as $C_A\ket{P_k} = \eta_k \ket{P_{k-1}}$ for $|\eta_k|=1$.
By successive redefinitions of states $\ket{P_k}$, we can assume $\eta_k=1$ for $k \neq 0 \mod N$.
Furthermore, the fact that the symmetry is $\bZ_N$ implies that $(C_A)^N=1$, and then we get $\prod_{k=1}^N \eta_k =1$.
From these facts, we can take all $\eta_k$ to be 1. Once we fix the phase factors of the states $\ket{P_k}$
in this way, there is no more freedom to change them except for the overall factor. Then the phase $e^{i\theta/N}$
appearing in \eqref{eq:transition} is well-defined, and hence $\theta$ has a physical meaning.
However, there is a freedom to change the definition of $C_A$ as $C'_A=e^{2\pi i/N}C_A$
and $\ket{P'_k}=e^{-2\pi i k/N} \ket{P_k}$. They still satisfy $C'_A\ket{P'_k} =  \ket{P'_{k-1}}$,
but we get the change $\theta \to \theta'=\theta + 2\pi$. Of course, this is the usual periodicity of $\theta$.}
Then we find that the basis $\ket{\el}$ diagonalizes the electric one-form center symmetry $C_A$
(hence the notation $\el$ for ``electric''):
\beq
C_A \ket{\el} = e^{\frac{2\pi i \el}{N}} \ket{\el} \;.
\eeq
Therefore, the different vacua has different charges under $C_A$. The $\el \in \bZ_N$ which determines the eigenvalues of $C_A$ is 
the 't~Hooft electric flux~\cite{tHooft:1979rtg,Witten:2000nv}.

We stress again that the center symmetry is still
preserved, because $C_A \ket{\el} \propto  \ket{\el} $ and hence they define the same ray in the Hilbert space (i.e., the same physical state).
So, the center symmetry is unbroken at each vacuum. By using \eqref{eq:wilAonP}, one can see that we can move between these vacua
by the action of the operator $\tr U_A$,
\beq
\frac{1}{N}\tr U_A \ket{\el} = \ket{\el+1} \;.
\eeq
This is consistent because $U_A$ is charged under $C_A$.

\section{Borel plane of QFTs and Linde's problem}\label{sec:BorelLinde}
Having constructed a setup where IR divergence is properly taken into account 
and the center symmetry ensured, we now have a setup where the resurgence program can be applicable, at least 
in principle. While the detailed analysis of such an analysis is left for future work,
let us here point out an important subtlety:
we demonstrate that there can exists singularities in Borel plane in QFT which depend on the shapes of manifolds.
Such manifold-dependent singularities could in some cases dominate over renormalons,
and hence dramatically change the singularity structure of the Borel plane of a compactified theory. 
This dependence is on the shapes of manifolds, so they survive even in the large volume limit.
This might suggest that there is no possibilities for resurgence directly on $\bR^4$ for Yang-Mills theory, and 
we always have to start from a compactified geometry without IR problems. 

\subsection{Quantum mechanics}

That we have manifold-dependent singularities in the Borel plane is a general phenomenon.
To explain this, let us start with 
1d quantum mechanics with target space $\CM$. 
The Lagrangian is
\beq
L_{\rm 1d}= \frac{1}{2g_{\rm QM}^2} h_{IJ} \partial_t \phi^I \partial_t \phi^J\;, \label{eq:QML}
\eeq
where $h_{IJ}$ is the metric on the target space $\CM$, and $g_{\rm QM}^2$ is the coupling constant (which is usually called $\hbar$). The canonical momentum is
\beq
\pi_I  = & \frac{\partial L_{\rm 1d}}{\partial (\partial_t \phi^I)}= \frac{1}{g_{\rm QM}^2} h_{IJ}  \partial_t \phi^J \nonumber 
\longrightarrow 
  -i \nabla_i \;,
\eeq
where $\nabla_I$ is the covariant derivative on the target space $\CM$ with respect to the metric $h_{IJ}$.
The Hamiltonian is therefore given by
\beq
H = \frac{g_{\rm QM}^2}{2} \Delta \;,
\eeq
where $\Delta=-\nabla^I \nabla_J$ is the Laplacian.

Let $\lambda_\ell$ be the $\ell$-th eingenvalue of the Laplacian $\Delta$, and let $n_\ell$ be the multiplicity of that eigenvalue. 
Then the partition function of the theory on $S^1$ with circumference $2 \pi R$ is given by
\beq
Z=\tr e^{- 2\pi R H } = \sum_{\ell = 0}^\infty n_\ell e^{- \pi R g_{\rm QM}^2 \lambda_\ell} \;.
\eeq

\paragraph{Asymptotic expansion.}
Now we restrict our attention to the case $\CM=\CP^1=S^2$ which has the unit radius. 
We want to perform perturbative expansion of $Z$ in terms of the coupling constant. 
In this case, we know that the eigenfunctions of the Laplacian $\Delta$ are given by spherical harmonics and
the eigenvalues and their degeneracies are given by
\beq
\lambda_\ell = \ell (\ell+1)\;, \qquad n_\ell=2\ell+1 \;,
\eeq
where $\ell=0$ is the ground state and $\ell \geq 1$ are excited states.
Now we can expand the partition function $Z$ by using the Euler-Maclaurin formula. If $f(x)$ is an analytic function, we have
asymptotic expansion given by
\beq
\sum_{k=0}^{K-1} f\left(k+\frac{1}{2}\right) = \int_0^K f(x)dx  - \sum_{m=1}^\infty \frac{(1-2^{-2m+1})B_{2m}}{(2m)!} [f^{(2m-1)}(0)-f^{(2m-1)}(K)]\;,
\eeq
where $f^{(2m-1)}$ means $(2m-1)$-th order differential of the function $f$.
The $B_{2m}$ are Bernoulli numbers and can be expressed as
\beq
B_{2m} &=(-1)^{m+1} \frac{2 (2m)!}{(2\pi)^{2m}} \zeta(2m)\;, \\
\zeta(s) &=\sum_{n=1}^\infty \frac{1}{n^s} \;.
\eeq
Note that $\zeta(s) \to 1$ as $s \to \infty$. Thus these formulas give the asymptotic behavior of $B_{2m}$.

Now we can rewrite the partition function of $\CP^1=S^2$ quantum mechanics.
We set
\beq
f(x) = xe^{- \pi R g_{\rm QM}^2 x^2 } \;.
\eeq
Then we get
\beq
e^{ -\frac{1}{4}\pi R g_{\rm QM}^2 } \cdot Z 
&= \sum_{\ell = 0}^\infty 2 \left( \ell +\frac{1}{2} \right)e^{- \pi R g_{\rm QM}^2 ( \ell+\frac{1}{2})^2 } \nonumber \\
&= 2\int_0^\infty f(x)dx  - 4\sum_{m=1}^\infty \frac{(1-2^{-2m+1})(-1)^{m+1} \zeta(2m)}{(2 \pi)^{2m}} f^{(2m-1)}(0)\;.
\eeq
One can compute
\beq
\int_0^\infty f(x)dx =\frac{1}{2\pi R g_{\rm QM}^2}\;,
\eeq
and 
\beq
f^{(2m-1)}(0) = (-1)^{m-1} \frac{(2m-1)!}{(m-1)!} (\pi R g_{\rm QM}^2)^{m-1}\;.
\eeq
Thus we get
\beq
e^{ -\frac{1}{4}\pi R g_{\rm QM}^2 } \cdot Z 
&=\frac{1}{\pi R g_{\rm QM}^2} - 4\sum_{m=1}^\infty \frac{(1-2^{-2m+1}) \zeta(2m)}{(2 \pi)^{2m}}  \frac{(2m-1)!}{(m-1)!} (\pi R g_{\rm QM}^2)^{m-1}\;.
\eeq

Let us study the asymptotic behavior. The coefficient grows as
\beq
\frac{(1-2^{-2m+1}) \zeta(2m)}{(2 \pi)^{2m}}  \frac{(2m-1)!}{(m-1)!} = C_m \frac{m!}{\pi^{2m}}\;,
\eeq
where $C_m \to 1/\sqrt{4\pi m}$. Therefore, a crude asymptotic behavior of $Z$ is given as
\beq
Z \sim & \sum_m m!  \left( \frac{R g_{\rm QM}^2 }{ \pi} \right)^{m} \;.\nonumber 
\eeq
This has a singularity on the positive real axis of the Borel plane with the associated ambiguity of order
\beq
\exp \left( -\frac{\pi}{R g_{\rm QM}^2 } \right)\;.
\eeq
Assuming that resurgence works,
this implies that the path integral of this quantum mechanics has unstable saddle points with the classical action $\pi /R g_{\rm QM}^2$. Interestingly, we can indeed find such a saddle point.

\paragraph{Classical saddle points.}
In the path integral with the Lagrangian \eqref{eq:QML}, there are saddle points
which are given by geodesics on the target space $\CM$. Let us consider the simple case 
$\CM = \CP^1=S^2$. Let $(\theta, \phi)$ be the polar coordinates of $\CM=S^2$.
Then, classical solutions of equations of motion are geodesics wrapping great circles of $S^2$.
For example, we have
\beq
(\theta, \phi) = \left(\frac{\pi}{2}\;,  \frac{nt}{R}\right) \qquad 0 \leq t \leq 2\pi R\;, \label{eq:clsol}
\eeq
where $n \in \bZ$. The action can be easily evaluated. The $n=0$ is the lowest stable solutions.
For $n= \pm 1$ it is given as
\beq
S_{\rm cl} = \int_0^{2\pi R} L dt  =  \frac{\pi }{Rg_{\rm QM}^2} \;.
\eeq
This is exactly as expected from the above analysis of the asymptotic expansion and the singularities of the Borel plane.
Notice that this action explicitly depends on the parameter $R$ which specify the manifold $S^1$.

\paragraph{Twist.}
We can also introduce a twist in the partition function above. Explicitly, we can impose the boundary condition such as
\beq
(\theta, \phi)_{(t+2\pi R)} = (\theta, \phi+\pi)_{(t)} \;.
\eeq
Then the classical solutions \eqref{eq:clsol} still exists, but now with the condition that $n \in \frac{1}{2} +\bZ$.
The stable lowest action solutions are $\theta =0, \pi$.
Therefore, the above solutions are unstable. The order of the classical action is the same up to numerical constants. 
The analysis of the asymptotic behavior of the perturbative expansion should also be straightforward, but we do not perform that explicitly.

\subsection{2d sigma models}
Having established that for 1d quantum mechanics the asymptotic behavior is reproduced from corresponding unstable saddle points,
we can easily see that there are singularities on the Borel plane of 2d sigma models. 

The Lagrangian for the 2d theory is
\beq
\CL_{\rm 2d} =\frac{1}{2g_{\rm 2d}^2} h_{IJ} \partial_i \phi^I \partial_i \phi^J \label{eq:QFTL}\;.
\eeq
Now let us consider partition function of this theory on the spacetime $\bT^2=S^1 \times S^1$, where
one $S^1$ has radius $R_1$ and the other $S^1$ has radius $R_2$; in this section we compactify the temporal direction 
of the 2d sigma models.
If $R_1 \ll R_2$, we may first dimensionally reduce the theory to the quantum mechanics discussed above.
The coupling constants are related as
\beq
g_{\rm QM}^2 = \frac{g_{\rm 2d}^2}{2\pi R_1}\;.
\eeq
The radiative corrections from the Kaluza-Klein modes are higher orders of $g_{\rm 2d}^2$.
Then, in the dimensionally reduced theory, we can use the above results about quantum mechanics.

Let us consider the case $\CM=\CP^1=S^2$. The classical action obtained above is given by
\beq
S_{\rm cl}  =  \frac{\pi }{R_2 g_{\rm QM}^2} = \frac{R_1}{R_2} \frac{2\pi^2}{g_{\rm 2d}^2}\;.\label{eq:shapedep}
\eeq
From this, we conclude that the Borel plane contains a singularity which depends on the shape $R_1/R_2$
of the manifold. Notice that when $R_1/R_2 \ll 1$,
this pole is very close to the origin than renormalons and instanton-anti-instanton pairs.

In the large volume limit
\beq
\frac{R_1}{R_2}=\text{fixed}\;, \qquad R_1R_2 \to \infty\;,
\eeq
we expect to recover the theory on flat space $\bR^2$.
In particular, we expect that
it is sensitive only to the volume $R_1R_2$ of spacetime, and not to the shape $R_1/R_2$ if the theory has a mass gap.\footnote{Instead, if the
theory is gapless in the large volume limit such as conformal field theories, it is natural that quantities such as partition functions depend on 
$R_1/R_2$.} 
Nevertheless, 
the above-mentioned singularity of the Borel plane with the corresponding action \eqref{eq:shapedep} still exists. This suggests that the structure of the Borel plane
is more complicated than is usually thought in this IR regulated situation.

On the other hand, if we take 
\beq
R_1=\text{fixed}\;, \qquad R_2 \to \infty\;,
\eeq
then the partition function gives the thermal free energy with temperature $T=(2\pi R_1)^{-1}$.
In this limit, the action \eqref{eq:shapedep} becomes extremely small and hence the singularity on the Borel plane merges into the origin.
This is a manifestation of the Linde's problem in thermal free energy in the present case of 2d $\CP^1$ sigma model. 

However, we stress that if we can complete the resurgence program of QFT,
we can still get a sensible answer. Let us demonstrate this point by using a toy function.
Consider
\beq
-\frac{1}{R_1R_2}\log Z_{\rm toy} = R_1^{-2}\int_0^1 dt \frac{\exp \left(-  \frac{R_1}{R_2} \frac{2\pi^2}{g_{\rm 2d}^2} t \right) }{\sqrt{1-t}}\;. \label{eq:toy}
\eeq
This has an asymptotic expansion given roughly as
\beq
-\frac{1}{R_1R_2}\log Z_{\rm toy} \sim R_1^{-2} \sum_m m! \left( \frac{R_2}{R_1} \frac{g_{\rm 2d}^2}{2\pi^2}  \right)\;.
\eeq
Naively from this perturbative expansion, 
it seems hopeless to get a sensible answer in the limit $R_2 \to \infty$. 
In the notation of the subsection~\ref{subsec:IR}, we identify $ \beta=T^{-1} \to 2\pi R_1$ and $m_{\rm IR} \to R_2^{-1}$,
so the IR cutoff $m_{\rm IR}$ goes to zero as $R_2 \to \infty$.
However, the integral in \eqref{eq:toy} makes perfect
sense in the limit $R_2 \to \infty$ with a finite value for the toy thermal free energy. 
Of course, the actual partition functions of QFT's are more complicated than the above toy function.
But the point here is that Linde's problem can in principle be overcome by resurgence.

\paragraph{Twist.}
Suppose that we introduce the twisted boundary condition on the $S^1$ with radius $R_2$.
If $R_1 \ll R_2$, the result is qualitatively similar to the one above; we have a singularity on the Borel plane which depends on $R_1/R_2$.
However, in the region $R_1 \gg R_2$, the theory has a mass gap of order $R_2^{-1}$ which is large,
and hence there is no IR divergences even in the limit $R_1 \to \infty$. Thus we expect that there is no problem in resurgence
in this limit. This is the situation studied in e.g., \cite{Dunne:2012ae,Dunne:2012zk,Dabrowski:2013kba,Bolognesi:2013tya,Misumi:2014jua,Misumi:2016fno,Fujimori:2016ljw} and also in this paper.

\subsection{Speculation on 4d Yang-Mills}
We would like to make speculative comments on the structure of the Borel plane and Linde's problem in 4d Yang-Mills.
By a compactification on small $\bT^2$ and considering only flat connections, we have seen that
the 4d $\SU(N)$ Yang-Mills is reduced to 2d $\CP^{N-1}$ at a qualitative level.\footnote{Quantitatively, the classical action 
of the 4d Yang-Mills gives a metric of $\CP^{N-1}$ which is different from the standard Fubini-Study metric. Also,
there are radiative corrections from W-bosons. It would be interesting to perform detailed computations on these points.}
Then, we expect that the 4d Yang-Mills encounters all the phenomena discussed above for $\CP^1$.

For example, we can find a classical solution of the Yang-Mills equation on a torus.
We consider the 4-manifold $\bT^4 =S^1_T \times S^1_A \times S^1_B \times S^1_C$ with the circumference of each circle given by
$L_T$, $L_A$, $L_B$ and $L_C$. We take the metric as
\beq
ds^2=L_T^2 dx_T^2+L_A^2 dx_A^2+L_B^2 dx_B^2+L_C^2 dx_C^2 \;,
\eeq
with each $x_{T,A,B,C}$ having periodicity $x_{T,A,B,C} \sim x_{T,A,B,C}+1$ .
Without twist, we can find a classical solution of the Yang-Mills equation as
\beq
iA=iA_\mu dx^\mu = \diag(a, -a,0,\cdots,0)\;,
\eeq
where 
\beq
a = 2\pi x_C dx_B\;.
\eeq
This solution is included in the Cartan of $\SU(N)$, and has a constant curvature
\beq
f = d a =2\pi dx_C \wedge dx_B\;.
\eeq
Thus it satisfies the Yang-Mills equations. Notice that this configuration $a = 2\pi x_C dx_B$
is similar to the classical solutions in $\CP^{N-1}$ given above via the relations between Yang-Mills and $\CP^{N-1}$ model
given in \eqref{eq:Wloop} and \eqref{eq:Atheta}.

With the twist on $S^1_B \times S^1_C$, we still have solutions such as
\beq
iA=iA_\mu dx^\mu = \diag(a, ,0,\cdots,0) - \frac{1}{N}\diag(a, a,\cdots,a)
\eeq
with the above $a$. 

These classical solutions have actions of order
\beq
S_{\rm cl} \sim  \frac{L_T L_A}{L_B L_C} \frac{8\pi^2}{g^2_{\rm YM}}\;,
\label{eq:S_LLLL}
\eeq
up to order 1 factor.
These actions explicitly depend on the shape of the 4-manifold even in the large volume limit.
We have not analyzed the asymptotic behavior of the perturbative expansion of the free energy, but
the existence of these unstable solutions suggests that the structure of the Borel plane is more complicated that is usually thought. 

If $L_T L_A \ll L_B L_C$, this classical action becomes arbitrary small. 
On the other hand, if  $L_T L_A \gg L_B L_C$, it depends on whether there is the twist or not.
Without the twist, we can have a similar solution $a =2\pi x_T dx_A$ with the action proportional to $(L_B L_C)/(L_T L_A)$.
So the situation is parallel to the above case. With twist, such a solution is forbidden basically because the twist 
generates a mass gap determined by $L_{B,C}$ and there is no IR divergences in the limit $L_{T,A} \to \infty$. 
Thus there is no classical saddle points with arbitrarily small action.

The value of the action \eqref{eq:S_LLLL} is very large in the parameter region $L_T \to \infty$ which we have studied in this paper, 
and hence the above unstable saddle point is unimportant for the resurgence analysis at zero temperature $L_T \to \infty$. 
By contrast, when we decompactify the spatial regions 
$L_{A,B,C} \to \infty$ with fixed ratios and constant finite $L_T$, then the action becomes arbitrarily small. 
This is a manifestation of Linde's problem of 4d Yang-Mills
and this problem should be overcome by resurgence as discussed above.


\appendix

\section{One-form global symmetry}\label{app:one_form}
Let us quickly review the concept of the one-form global symmetry~\cite{Kapustin:2014gua,Gaiotto:2014kfa}.
The usual (i.e., 0-form) symmetries act on local operators, while 1-form symmetries act on line operators such as Wilson loop operators. More explicitly, they can be described as follows. The usual 0-form symmetry for a continuous symmetry 
can be described by picking up a codimension-1
submanifold $Y$ inside a spacetime $X$. By integrating a current $j =j_\mu dx^\mu$ on $Y$ we get a charge as $Q(Y)=\int_Y * j$.
For discrete symmetries, there is no current operator, but we can still consider charge operators $Q(Y)$ defined on $Y$.
Conservation of the charge is translated to the statement that this operator $Q(Y)$ only depends on the topology of $Y$ outside the locus where
other operators are inserted. If we cross a local operator
across $Y$, then that operator is transformed by $Q(Y)$. We can do similar things for $p$-form symmetries.
Pick up a codimension-$(p+1)$ submanifold $Y$ and define a charge operator $Q(Y)$. 
Operators charged under them are extended in $p$ dimensional
manifold $Z$. 
We denote that operator as $O(X)$.
If the $Z$ is crossed across $Y$, the operator under consideration
is changed by the charge operator $Q(Y)$. 

For example, if $Z \cong \bR^{p}$ is a flat subspace inside the total spacetime $X=\bR^d$, and $Y \cong S^{d-p-1}$ is a sphere surrounding $Z$
where $d = \dim X$, then we get
\beq
Q(Y) \, O(Z) =q\, O(Z)\;,
\eeq
where $q$ is the charge of the operator $O$ under the generator $Q$ of the symmetry.

If we consider the case $p=1$, $d=4$ and $O(Z)=\tr U(Z)$ is the Wilson loop operator, there is a 1-form $\bZ_N$ symmetry $C$ such that\footnote{$C(Y)$ here is a Gukov-Witten surface operator \cite{Gukov:2008sn,Gukov:2006jk}.}
\beq
C(Y)\tr U(Z) = e^{\frac{2\pi i L(Y,Z)}{N}} \tr U(Z)\;,
\eeq
where $L(Y,Z)$ is the linking number of $Y$ and $Z$. For example, in the case discussed above, 
namely $Z \cong \bR^{p}$ and $Y \cong S^{d-p-1}$ is a sphere surrounding $Z$, we get $L(Y,Z)=1$.

\acknowledgments
We would like to thank Kentaro Hori for helpful discussions on the relation between Yang-Mills and $\CP^{N-1}$,
and Taizan Watari for lectures on algebraic geometry. The workshop ``Resurgence at Kavli IPMU'' was inspiring for the initiation of this research.
MY and KY are supported in part by the WPI Research Center Initiative (MEXT, Japan). MY is also supported by JSPS Program for Advancing Strategic International Networks to Accelerate the Circulation of Talented Researchers, by JSPS KAKENHI (15K17634), and by JSPS-NRF Joint Research Project. He would like to thank Center for the Fundamental Laws of Nature, Harvard University, for hospitality where part of this work was done. The work of KY is also supported by JSPS KAKENHI Grant-in-Aid (17K14265).


\bibliographystyle{JHEP}
\bibliography{ref}

\providecommand{\href}[2]{#2}\begingroup\raggedright\begin{thebibliography}{10}

\bibitem{Costello_Book}
K.~Costello, {\em Renormalization and effective field theory}, vol.~170 of {\em
  Mathematical Surveys and Monographs}.
\newblock American Mathematical Society, Providence, RI, 2011.

\bibitem{tHooft:1977xjm}
G.~'t~Hooft, {\it {Can We Make Sense Out of Quantum Chromodynamics?}},  {\em
  Subnucl. Ser.} {\bf 15} (1979) 943.

\bibitem{Ecalle1}
J.~\'Ecalle, {\em Les fonctions r\'esurgentes. {T}ome {I}}, vol.~5 of {\em
  Publications Math\'ematiques d'Orsay 81 [Mathematical Publications of Orsay
  81]}.
\newblock Universit\'e de Paris-Sud, D\'epartement de Math\'ematique, Orsay,
  1981.
\newblock Les alg\`ebres de fonctions r\'esurgentes. [The algebras of resurgent
  functions], With an English foreword.

\bibitem{Ecalle2}
J.~\'Ecalle, {\em Les fonctions r\'esurgentes. {T}ome {II}}, vol.~6 of {\em
  Publications Math\'ematiques d'Orsay 81 [Mathematical Publications of Orsay
  81]}.
\newblock Universit\'e de Paris-Sud, D\'epartement de Math\'ematique, Orsay,
  1981.
\newblock Les fonctions r\'esurgentes appliqu\'ees \`a l'it\'eration.
  [Resurgent functions applied to iteration].

\bibitem{Costin_Book}
O.~Costin, {\em Asymptotics and {B}orel summability}, vol.~141 of {\em Chapman
  \& Hall/CRC Monographs and Surveys in Pure and Applied Mathematics}.
\newblock CRC Press, Boca Raton, FL, 2009.

\bibitem{Sauzin_Book}
C.~Mitschi and D.~Sauzin, {\em Divergent series, summability and resurgence.
  {I}}, vol.~2153 of {\em Lecture Notes in Mathematics}.
\newblock Springer, [Cham], 2016.
\newblock Monodromy and resurgence, With a foreword by Jean-Pierre Ramis and a
  preface by \'Eric Delabaere, Mich\`ele Loday-Richaud, Claude Mitschi and
  David Sauzin.

\bibitem{Bogomolny:1980ur}
E.~B. Bogomolny, {\it {CALCULATION OF INSTANTON - ANTI-INSTANTON CONTRIBUTIONS
  IN QUANTUM MECHANICS}},  {\em Phys. Lett.} {\bf B91} (1980) 431--435.

\bibitem{ZinnJustin:1981dx}
J.~Zinn-Justin, {\it {Multi - Instanton Contributions in Quantum Mechanics}},
  {\em Nucl. Phys.} {\bf B192} (1981) 125--140.

\bibitem{Pasquetti:2009jg}
S.~Pasquetti and R.~Schiappa, {\it {Borel and Stokes Nonperturbative Phenomena
  in Topological String Theory and c=1 Matrix Models}},  {\em Annales Henri
  Poincare} {\bf 11} (2010) 351--431,
  [\href{http://arxiv.org/abs/0907.4082}{{\tt arXiv:0907.4082}}].

\bibitem{Aniceto:2014hoa}
I.~Aniceto, J.~G. Russo, and R.~Schiappa, {\it {Resurgent Analysis of
  Localizable Observables in Supersymmetric Gauge Theories}},  {\em JHEP} {\bf
  03} (2015) 172, [\href{http://arxiv.org/abs/1410.5834}{{\tt
  arXiv:1410.5834}}].

\bibitem{Honda:2016mvg}
M.~Honda, {\it {Borel Summability of Perturbative Series in 4D $N=2$ and 5D
  $N$=1 Supersymmetric Theories}},  {\em Phys. Rev. Lett.} {\bf 116} (2016),
  no.~21 211601, [\href{http://arxiv.org/abs/1603.06207}{{\tt
  arXiv:1603.06207}}].

\bibitem{Honda:2016vmv}
M.~Honda, {\it {How to resum perturbative series in 3d N=2 Chern-Simons matter
  theories}},  {\em Phys. Rev.} {\bf D94} (2016), no.~2 025039,
  [\href{http://arxiv.org/abs/1604.08653}{{\tt arXiv:1604.08653}}].

\bibitem{Dunne:2016jsr}
G.~V. Dunne and M.~Unsal, {\it {Deconstructing zero: resurgence, supersymmetry
  and complex saddles}},  {\em JHEP} {\bf 12} (2016) 002,
  [\href{http://arxiv.org/abs/1609.05770}{{\tt arXiv:1609.05770}}].

\bibitem{Fujimori:2017oab}
T.~Fujimori, S.~Kamata, T.~Misumi, M.~Nitta, and N.~Sakai, {\it {Exact
  resurgent trans-series and multibion contributions to all orders}},  {\em
  Phys. Rev.} {\bf D95} (2017), no.~10 105001,
  [\href{http://arxiv.org/abs/1702.00589}{{\tt arXiv:1702.00589}}].

\bibitem{Unsal:2007vu}
M.~Unsal, {\it {Abelian Duality, Confinement, and Chiral Symmetry Breaking in
  QCD(Adj)}},  {\em Phys. Rev. Lett.} {\bf 100} (2008) 032005,
  [\href{http://arxiv.org/abs/0708.1772}{{\tt arXiv:0708.1772}}].

\bibitem{Kovtun:2007py}
P.~Kovtun, M.~Unsal, and L.~G. Yaffe, {\it {Volume independence in large N(c)
  QCD-like gauge theories}},  {\em JHEP} {\bf 06} (2007) 019,
  [\href{http://arxiv.org/abs/hep-th/0702021}{{\tt hep-th/0702021}}].

\bibitem{Unsal:2007jx}
M.~Unsal, {\it {Magnetic bion condensation: A New mechanism of confinement and
  mass gap in four dimensions}},  {\em Phys. Rev.} {\bf D80} (2009) 065001,
  [\href{http://arxiv.org/abs/0709.3269}{{\tt arXiv:0709.3269}}].

\bibitem{Unsal:2008ch}
M.~Unsal and L.~G. Yaffe, {\it {Center-stabilized Yang-Mills theory:
  Confinement and large N volume independence}},  {\em Phys. Rev.} {\bf D78}
  (2008) 065035, [\href{http://arxiv.org/abs/0803.0344}{{\tt
  arXiv:0803.0344}}].

\bibitem{Shifman:2009tp}
M.~Shifman and M.~Unsal, {\it {Multiflavor QCD* on R(3) $\times$ $S^1$:
  Studying Transition from Abelian to Non-Abelian Confinement}},  {\em Phys.
  Lett.} {\bf B681} (2009) 491--494,
  [\href{http://arxiv.org/abs/0901.3743}{{\tt arXiv:0901.3743}}].

\bibitem{Shifman:2008ja}
M.~Shifman and M.~Unsal, {\it {QCD-Like Theories on R(3) $\times$ $S^1$: a
  Smooth Journey from Small to Large R($S^1$) with Double-Trace Deformations}},
   {\em Phys. Rev.} {\bf D78} (2008) 065004,
  [\href{http://arxiv.org/abs/0802.1232}{{\tt arXiv:0802.1232}}].

\bibitem{Argyres:2012ka}
P.~C. Argyres and M.~Unsal, {\it {The semi-classical expansion and resurgence
  in gauge theories: new perturbative, instanton, bion, and renormalon
  effects}},  {\em JHEP} {\bf 08} (2012) 063,
  [\href{http://arxiv.org/abs/1206.1890}{{\tt arXiv:1206.1890}}].

\bibitem{Argyres:2012vv}
P.~Argyres and M.~Unsal, {\it {A semiclassical realization of infrared
  renormalons}},  {\em Phys. Rev. Lett.} {\bf 109} (2012) 121601,
  [\href{http://arxiv.org/abs/1204.1661}{{\tt arXiv:1204.1661}}].

\bibitem{Dunne:2012ae}
G.~V. Dunne and M.~Unsal, {\it {Resurgence and Trans-series in Quantum Field
  Theory: The CP(N-1) Model}},  {\em JHEP} {\bf 11} (2012) 170,
  [\href{http://arxiv.org/abs/1210.2423}{{\tt arXiv:1210.2423}}].

\bibitem{Dunne:2012zk}
G.~V. Dunne and M.~Unsal, {\it {Continuity and Resurgence: towards a continuum
  definition of the $\mathbb{CP}$(N-1) model}},  {\em Phys. Rev.} {\bf D87}
  (2013) 025015, [\href{http://arxiv.org/abs/1210.3646}{{\tt
  arXiv:1210.3646}}].

\bibitem{Poppitz:2012sw}
E.~Poppitz, T.~Schafer, and M.~Unsal, {\it {Continuity, Deconfinement, and
  (Super) Yang-Mills Theory}},  {\em JHEP} {\bf 10} (2012) 115,
  [\href{http://arxiv.org/abs/1205.0290}{{\tt arXiv:1205.0290}}].

\bibitem{Anber:2013doa}
M.~M. Anber, S.~Collier, E.~Poppitz, S.~Strimas-Mackey, and B.~Teeple, {\it
  {Deconfinement in $\mathcal{N}=1$ Super Yang-Mills Theory on $\mathbb{R}^3
  \times \mathbb{S}^1$ via Dual-Coulomb Gas and "Affine" Xy-Model}},  {\em
  JHEP} {\bf 11} (2013) 142, [\href{http://arxiv.org/abs/1310.3522}{{\tt
  arXiv:1310.3522}}].

\bibitem{Cherman:2013yfa}
A.~Cherman, D.~Dorigoni, G.~V. Dunne, and M.~Unsal, {\it {Resurgence in Quantum
  Field Theory: Nonperturbative Effects in the Principal Chiral Model}},  {\em
  Phys. Rev. Lett.} {\bf 112} (2014) 021601,
  [\href{http://arxiv.org/abs/1308.0127}{{\tt arXiv:1308.0127}}].

\bibitem{Cherman:2014ofa}
A.~Cherman, D.~Dorigoni, and M.~Unsal, {\it {Decoding Perturbation Theory Using
  Resurgence: Stokes Phenomena, New Saddle Points and Lefschetz Thimbles}},
  {\em JHEP} {\bf 10} (2015) 056, [\href{http://arxiv.org/abs/1403.1277}{{\tt
  arXiv:1403.1277}}].

\bibitem{Dunne:2015ywa}
G.~V. Dunne and M.~Unsal, {\it {Resurgence and Dynamics of $O(N)$ and
  Grassmannian Sigma Models}},  {\em JHEP} {\bf 09} (2015) 199,
  [\href{http://arxiv.org/abs/1505.07803}{{\tt arXiv:1505.07803}}].

\bibitem{Misumi:2014bsa}
T.~Misumi, M.~Nitta, and N.~Sakai, {\it {Classifying Bions in Grassmann Sigma
  Models and Non-Abelian Gauge Theories by D-Branes}},  {\em PTEP} {\bf 2015}
  (2015) 033B02, [\href{http://arxiv.org/abs/1409.3444}{{\tt
  arXiv:1409.3444}}].

\bibitem{Misumi:2014jua}
T.~Misumi, M.~Nitta, and N.~Sakai, {\it {Neutral bions in the ${\mathbb
  C}P^{N-1}$ model}},  {\em JHEP} {\bf 06} (2014) 164,
  [\href{http://arxiv.org/abs/1404.7225}{{\tt arXiv:1404.7225}}].

\bibitem{Dunne:2016nmc}
G.~V. Dunne and M.~Unsal, {\it {New Nonperturbative Methods in Quantum Field
  Theory: from Large-N Orbifold Equivalence to Bions and Resurgence}},  {\em
  Ann. Rev. Nucl. Part. Sci.} {\bf 66} (2016) 245--272,
  [\href{http://arxiv.org/abs/1601.03414}{{\tt arXiv:1601.03414}}].

\bibitem{Cherman:2016hcd}
A.~Cherman, T.~Schafer, and M.~Unsal, {\it {Chiral Lagrangian from Duality and
  Monopole Operators in Compactified QCD}},  {\em Phys. Rev. Lett.} {\bf 117}
  (2016), no.~8 081601, [\href{http://arxiv.org/abs/1604.06108}{{\tt
  arXiv:1604.06108}}].

\bibitem{Sulejmanpasic:2016llc}
T.~Sulejmanpasic, {\it {Global Symmetries, Volume Independence, and Continuity
  in Quantum Field Theories}},  {\em Phys. Rev. Lett.} {\bf 118} (2017), no.~1
  011601, [\href{http://arxiv.org/abs/1610.04009}{{\tt arXiv:1610.04009}}].

\bibitem{GonzalezArroyo:1982ub}
A.~Gonzalez-Arroyo and M.~Okawa, {\it {A Twisted Model for Large $N$ Lattice
  Gauge Theory}},  {\em Phys. Lett.} {\bf B120} (1983) 174--178.

\bibitem{GonzalezArroyo:1982hz}
A.~Gonzalez-Arroyo and M.~Okawa, {\it {The Twisted Eguchi-Kawai Model: A
  Reduced Model for Large N Lattice Gauge Theory}},  {\em Phys. Rev.} {\bf D27}
  (1983) 2397.

\bibitem{GonzalezArroyo:2010ss}
A.~Gonzalez-Arroyo and M.~Okawa, {\it {Large $N$ reduction with the Twisted
  Eguchi-Kawai model}},  {\em JHEP} {\bf 07} (2010) 043,
  [\href{http://arxiv.org/abs/1005.1981}{{\tt arXiv:1005.1981}}].

\bibitem{Anber:2014sda}
M.~M. Anber and T.~Sulejmanpasic, {\it {The renormalon diagram in gauge
  theories on $ {\mathrm{\mathbb{R}}}3\times {\mathbb{S}}1 $}},  {\em JHEP}
  {\bf 01} (2015) 139, [\href{http://arxiv.org/abs/1410.0121}{{\tt
  arXiv:1410.0121}}].

\bibitem{tHooft:1976snw}
G.~'t~Hooft, {\it {Computation of the Quantum Effects Due to a Four-Dimensional
  Pseudoparticle}},  {\em Phys. Rev.} {\bf D14} (1976) 3432--3450. [Erratum:
  Phys. Rev.D18,2199(1978)].

\bibitem{Vafa:1984xg}
C.~Vafa and E.~Witten, {\it {Parity Conservation in QCD}},  {\em Phys. Rev.
  Lett.} {\bf 53} (1984) 535.

\bibitem{Witten:1980sp}
E.~Witten, {\it {Large N Chiral Dynamics}},  {\em Annals Phys.} {\bf 128}
  (1980) 363.

\bibitem{Witten:1998uka}
E.~Witten, {\it {Theta dependence in the large N limit of four-dimensional
  gauge theories}},  {\em Phys. Rev. Lett.} {\bf 81} (1998) 2862--2865,
  [\href{http://arxiv.org/abs/hep-th/9807109}{{\tt hep-th/9807109}}].

\bibitem{Giusti:2007tu}
L.~Giusti, S.~Petrarca, and B.~Taglienti, {\it {Theta dependence of the vacuum
  energy in the SU(3) gauge theory from the lattice}},  {\em Phys. Rev.} {\bf
  D76} (2007) 094510, [\href{http://arxiv.org/abs/0705.2352}{{\tt
  arXiv:0705.2352}}].

\bibitem{Unsal:2012zj}
M.~Unsal, {\it {Theta dependence, sign problems and topological interference}},
   {\em Phys. Rev.} {\bf D86} (2012) 105012,
  [\href{http://arxiv.org/abs/1201.6426}{{\tt arXiv:1201.6426}}].

\bibitem{Yonekura:2014oja}
K.~Yonekura, {\it {Notes on natural inflation}},  {\em JCAP} {\bf 1410} (2014),
  no.~10 054, [\href{http://arxiv.org/abs/1405.0734}{{\tt arXiv:1405.0734}}].

\bibitem{Gaiotto:2017yup}
D.~Gaiotto, A.~Kapustin, Z.~Komargodski, and N.~Seiberg, {\it {Theta, Time
  Reversal, and Temperature}},  \href{http://arxiv.org/abs/1703.00501}{{\tt
  arXiv:1703.00501}}.

\bibitem{Linde:1980ts}
A.~D. Linde, {\it {Infrared Problem in Thermodynamics of the Yang-Mills Gas}},
  {\em Phys. Lett.} {\bf 96B} (1980) 289--292.

\bibitem{Gaiotto:2014kfa}
D.~Gaiotto, A.~Kapustin, N.~Seiberg, and B.~Willett, {\it {Generalized Global
  Symmetries}},  {\em JHEP} {\bf 02} (2015) 172,
  [\href{http://arxiv.org/abs/1412.5148}{{\tt arXiv:1412.5148}}].

\bibitem{Looijenga1}
E.~Looijenga, {\it {Root Systems And Elliptic Curves}},  {\em Invent. Math.}
  {\bf {\bf 38}} (1977).

\bibitem{Looijenga2}
E.~Looijenga, {\it {Invariant Theory For Generalized Root Sytems}},  {\em
  Invent. Math.} {\bf {\bf 61}} (1980).

\bibitem{Friedman:1997yq}
R.~Friedman, J.~Morgan, and E.~Witten, {\it {Vector bundles and F theory}},
  {\em Commun. Math. Phys.} {\bf 187} (1997) 679--743,
  [\href{http://arxiv.org/abs/hep-th/9701162}{{\tt hep-th/9701162}}].

\bibitem{GH}
P.~Griffiths and J.~Harris, {\em Principles of algebraic geometry}.
\newblock Wiley-Interscience [John Wiley \&\ Sons], New York, 1978.
\newblock Pure and Applied Mathematics.

\bibitem{Atiyah:1984tk}
M.~F. Atiyah, {\it {INSTANTONS IN TWO-DIMENSIONS AND FOUR-DIMENSIONS}},  {\em
  Commun. Math. Phys.} {\bf 93} (1984) 437--451.

\bibitem{Kapustin:2006pk}
A.~Kapustin and E.~Witten, {\it {Electric-Magnetic Duality And The Geometric
  Langlands Program}},  {\em Commun. Num. Theor. Phys.} {\bf 1} (2007) 1--236,
  [\href{http://arxiv.org/abs/hep-th/0604151}{{\tt hep-th/0604151}}].

\bibitem{Atiyah:1978wi}
M.~F. Atiyah, N.~J. Hitchin, and I.~M. Singer, {\it {Selfduality in
  Four-Dimensional Riemannian Geometry}},  {\em Proc. Roy. Soc. Lond.} {\bf
  A362} (1978) 425--461.

\bibitem{Eto:2004rz}
M.~Eto, Y.~Isozumi, M.~Nitta, K.~Ohashi, and N.~Sakai, {\it {Instantons in the
  Higgs phase}},  {\em Phys. Rev.} {\bf D72} (2005) 025011,
  [\href{http://arxiv.org/abs/hep-th/0412048}{{\tt hep-th/0412048}}].

\bibitem{Eto:2006mz}
M.~Eto, T.~Fujimori, Y.~Isozumi, M.~Nitta, K.~Ohashi, K.~Ohta, and N.~Sakai,
  {\it {Non-Abelian vortices on cylinder: Duality between vortices and walls}},
   {\em Phys. Rev.} {\bf D73} (2006) 085008,
  [\href{http://arxiv.org/abs/hep-th/0601181}{{\tt hep-th/0601181}}].

\bibitem{Bruckmann:2007zh}
F.~Bruckmann, {\it {Instanton constituents in the O(3) model at finite
  temperature}},  {\em Phys. Rev. Lett.} {\bf 100} (2008) 051602,
  [\href{http://arxiv.org/abs/0707.0775}{{\tt arXiv:0707.0775}}].

\bibitem{Brendel:2009mp}
W.~Brendel, F.~Bruckmann, L.~Janssen, A.~Wipf, and C.~Wozar, {\it {Instanton
  constituents and fermionic zero modes in twisted CP**n models}},  {\em Phys.
  Lett.} {\bf B676} (2009) 116--125,
  [\href{http://arxiv.org/abs/0902.2328}{{\tt arXiv:0902.2328}}].

\bibitem{Harland:2009mf}
D.~Harland, {\it {Kinks, chains, and loop groups in the CP**n sigma models}},
  {\em J. Math. Phys.} {\bf 50} (2009) 122902,
  [\href{http://arxiv.org/abs/0902.2303}{{\tt arXiv:0902.2303}}].

\bibitem{Ohmori:2015pua}
K.~Ohmori, H.~Shimizu, Y.~Tachikawa, and K.~Yonekura, {\it {6d
  $\mathcal{N}=(1,0)$ theories on $T^2$ and class S theories: Part I}},  {\em
  JHEP} {\bf 07} (2015) 014, [\href{http://arxiv.org/abs/1503.06217}{{\tt
  arXiv:1503.06217}}].

\bibitem{Witten:1979ey}
E.~Witten, {\it {Dyons of Charge e theta/2 pi}},  {\em Phys. Lett.} {\bf B86}
  (1979) 283--287.

\bibitem{tHooft:1979rtg}
G.~'t~Hooft, {\it {A Property of Electric and Magnetic Flux in Nonabelian Gauge
  Theories}},  {\em Nucl. Phys.} {\bf B153} (1979) 141--160.

\bibitem{Witten:2000nv}
E.~Witten, {\it {Supersymmetric index in four-dimensional gauge theories}},
  {\em Adv. Theor. Math. Phys.} {\bf 5} (2002) 841--907,
  [\href{http://arxiv.org/abs/hep-th/0006010}{{\tt hep-th/0006010}}].

\bibitem{Dabrowski:2013kba}
R.~Dabrowski and G.~V. Dunne, {\it {Fractionalized Non-Self-Dual Solutions in
  the CP(N-1) Model}},  {\em Phys. Rev.} {\bf D88} (2013), no.~2 025020,
  [\href{http://arxiv.org/abs/1306.0921}{{\tt arXiv:1306.0921}}].

\bibitem{Bolognesi:2013tya}
S.~Bolognesi and W.~Zakrzewski, {\it {Clustering and decomposition for non-BPS
  solutions of the $\mathbb{CP}^{N-1}$ models}},  {\em Phys. Rev.} {\bf D89}
  (2014), no.~6 065013, [\href{http://arxiv.org/abs/1310.8247}{{\tt
  arXiv:1310.8247}}].

\bibitem{Misumi:2016fno}
T.~Misumi, M.~Nitta, and N.~Sakai, {\it {Non-BPS exact solutions and their
  relation to bions in ${\mathbb C}P^{N-1}$ models}},  {\em JHEP} {\bf 05}
  (2016) 057, [\href{http://arxiv.org/abs/1604.00839}{{\tt arXiv:1604.00839}}].

\bibitem{Fujimori:2016ljw}
T.~Fujimori, S.~Kamata, T.~Misumi, M.~Nitta, and N.~Sakai, {\it
  {Nonperturbative contributions from complexified solutions in
  $\mathbb{C}P^{N-1}$models}},  {\em Phys. Rev.} {\bf D94} (2016), no.~10
  105002, [\href{http://arxiv.org/abs/1607.04205}{{\tt arXiv:1607.04205}}].

\bibitem{Kapustin:2014gua}
A.~Kapustin and N.~Seiberg, {\it {Coupling a QFT to a TQFT and Duality}},  {\em
  JHEP} {\bf 04} (2014) 001, [\href{http://arxiv.org/abs/1401.0740}{{\tt
  arXiv:1401.0740}}].

\bibitem{Gukov:2008sn}
S.~Gukov and E.~Witten, {\it {Rigid Surface Operators}},  {\em Adv. Theor.
  Math. Phys.} {\bf 14} (2010), no.~1 87--178,
  [\href{http://arxiv.org/abs/0804.1561}{{\tt arXiv:0804.1561}}].

\bibitem{Gukov:2006jk}
S.~Gukov and E.~Witten, {\it {Gauge Theory, Ramification, And The Geometric
  Langlands Program}},  \href{http://arxiv.org/abs/hep-th/0612073}{{\tt
  hep-th/0612073}}.

\end{thebibliography}\endgroup

\end{document}